\documentclass[11pt]{article}
\usepackage{fullpage}
\usepackage{epic,eepic,epsfig,graphics,amsmath,color,fullpage}
\usepackage{amsmath}
\usepackage{latexsym}
\usepackage{times}
\usepackage{graphicx}
\graphicspath{ {./images/} }
\usepackage{hyperref}
\usepackage{comment}
\renewcommand{\baselinestretch}{1}
\usepackage{tikz}
\usepackage{textcomp}
\usepackage[maxnames=99, style=ieee]{biblatex}
\usepackage{booktabs}
\usepackage{tabularx}
\usepackage{import}
\usepackage{verbatim}
\usetikzlibrary{backgrounds,shapes,arrows,positioning,calc,fit,shadows,trees,mindmap,backgrounds}
\usepgflibrary{decorations.markings}
\usepackage{dtk-logos}
\usepackage{pgfplots}
\usepackage{subfig}
\usepackage[edges]{forest}
\usepackage{amsfonts}
\usepackage{multirow}
\usepackage{graphicx}
\usepackage[linesnumbered,ruled]{algorithm2e}
\usepackage{algorithmic}
\usetikzlibrary{arrows.meta,
                quotes,
                shapes.geometric}
\usetikzlibrary{shapes,arrows}
\usepackage{float}
\usepackage{lscape}
\usetikzlibrary{decorations.text}
\definecolor{mypink1}{rgb}{0.858, 0.188, 0.478}

\pgfplotsset{width=6cm,compat=1.5}
  \def\angle{0}
\def\radius{3}

\def\cyclelist{{"teal!65","blue!40","teal!20","cyan!40"}}
\newcount\cyclecount \cyclecount=-1
\newcount\ind \ind=-1

\tikzset{
  basic/.style  = {draw, text width=2cm, drop shadow, font=\sffamily, rectangle},
  root/.style   = {basic, rounded corners=8pt, thin, align=center,
                   fill=blue!70},
  level 2/.style = {basic, rounded corners=8pt, thin,align=center, fill=blue!50,
                   text width=3cm},
  level 3/.style = {basic, thin, align=left, fill=blue!20, text width=1cm}
}

\usetikzlibrary{arrows}
\usetikzlibrary{shapes}

\usepackage{biblatex}
\begin{document}

\title{\bf Byzantine Fault Tolerance in Distributed Machine Learning : a Survey 
}

\author{Djamila Bouhata$^{1}$~~
        Hamouma Moumen$^{1}$\footnote{Correspondence: Prof. Hamouma Moumen. Email: hamouma.moumen@univ-batna2.dz.}~~
        Jocelyn Ahmed Mazari$^{2,3}$~~
        Ahcène Bounceur$^{4}$~~\\~\\
$^{1}$ University of Batna 2, 05000 Batna, Algeria \\
$^{2}$ University of Sorbonne , CNRS, ISIR, F-75005 Paris, France\\ 
$^{3}$ Extrality, 75002 Paris, France\\
$^{4}$ Lab-STICC UMR CNRS, University of Western Brittany, UBO, Brest, 6285, France.\\
{\small {\tt \{dj.bouhata|hamouma.moumen\}@univ-batna2.dz}}\\
{\small {\tt \{ahmed\}@extrality.ai}}\\
{\small {\tt \{Ahcene.Bounceur\}@univ-brest.fr}}
}

\date{}
\maketitle
\renewcommand{\baselinestretch}{1,0}

\begin{abstract}
\label{abst}
Byzantine Fault Tolerance (BFT) is one of the most challenging problems in Distributed Machine Learning (DML), defined as the resilience of a fault-tolerant system in the presence of malicious components. Byzantine failures are still difficult to deal with due to their unrestricted nature, which results in the possibility of generating arbitrary data. Significant research efforts are constantly being made to implement BFT in DML. Some recent studies have considered various BFT approaches in DML. However, some aspects are limited, such as the few approaches analyzed, and there is no classification of the techniques used in the studied approaches. In this paper, we present a survey of recent work surrounding BFT in DML, mainly in first-order optimization methods, especially Stochastic Gradient Descent (SGD). We highlight key techniques as well as fundamental approaches. We provide an illustrative description of the techniques used in BFT in DML, with a proposed classification of BFT approaches in the context of their fundamental techniques. This classification is established on specific criteria such as communication process, optimization method, and topology setting, which characterize future work methods addressing open challenges. 

~\\~\\
\noindent
{\bf Keywords}: Byzantine fault tolerance, distributed machine learning, stochastic gradient descent, filtering scheme, coding scheme, Blockchain.
\end{abstract}

\maketitle

%------------------------Section1--------------------------------
\section{Introduction}
\label{Intro}
\begin{figure}[t]
    \centering
    \includegraphics[width=150mm,scale=1]{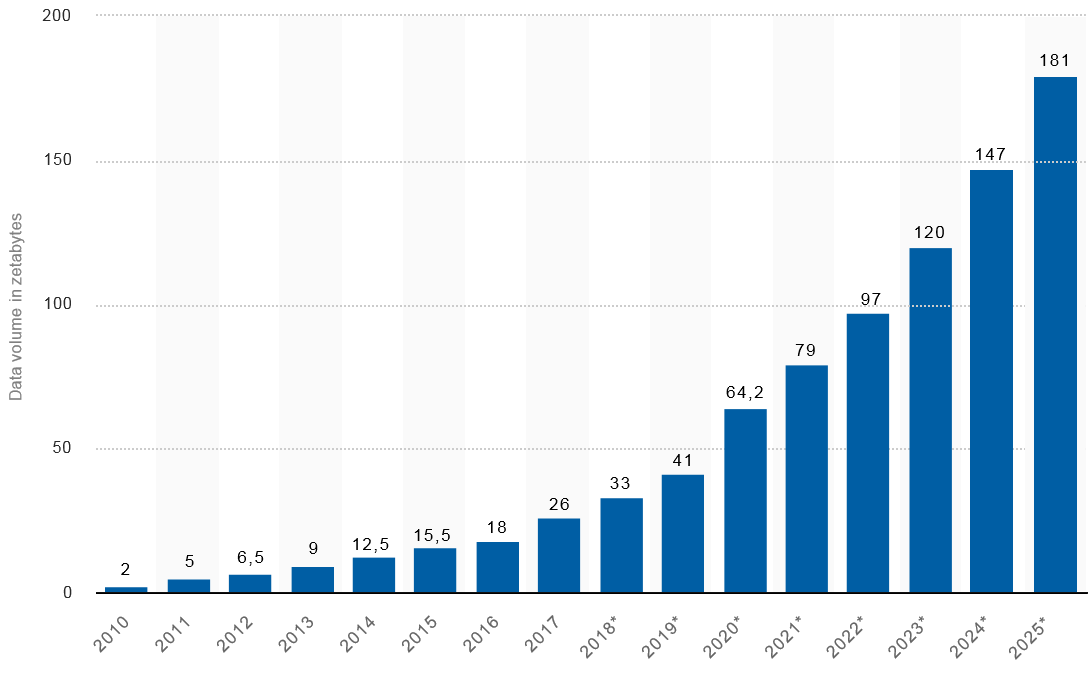}
    \caption{Growth of data volume in zetabytes (1 zetabyte = $10^{21}$ bytes = 1000000000 terabytes) from 2010 to 2025 \cite{statista-2022}. The star (*) in the chart reflects statista's prediction of rising global data from 2021 to 2025, based on reached 2020 figure with a 5-year compound annual growth rate (CAGR). In comparison, the International Data Corporation (IDC) provides the prediction figures from 2018 to 2020.}

    \label{fig:GrowthData}
\end{figure}

Over the last two decades, data representation has evolved significantly. Data from different receiving sources, such as machine log data\footnote{Machine log data is machine-generated data like event logs, server data, and application logs.} and social media, need to be processed, analyzed, and stored. With the growth in the use of information systems, data volumes have increased dramatically (see Fig.~\ref{fig:GrowthData}). One of the most promising approaches to analyzing big data is machine learning (ML), which can process this data efficiently. However, due to the ``over-fitting and over-parameterized model''\footnote{For large and complex neural networks, an algorithm may take a long time to run, indicating over-learning (or over-fitting); this shows that while the learning system excels at fitting its training data, it cannot generalize well to new or unseen data. On the other hand, the over-parameterized model represents the number of parameters when a model can contain more than the number of parameters estimated from the data.} in some cases, such as neural networks \cite{panchal2011determination}, the challenging tasks presented in \cite{yang2018end,leopold2018identifying} as well as the centralized solutions suffer from the drastic growth of the data when it is impossible to store it on single machines \cite{raicu2006astroportal}. These difficulties make solutions from traditional ML algorithms not feasible \cite{verbraeken2020survey}. In addition, training large-scale neural network models on a single machine is computationally challenging. Therefore, using distributed systems allows optimization algorithms to be more efficient in terms of computational and communication costs while maintaining similar accuracy to individual machines.
\subsection{Domain of the survey}
\begin{figure}[t]
    \centering
    \includegraphics[width=\linewidth]{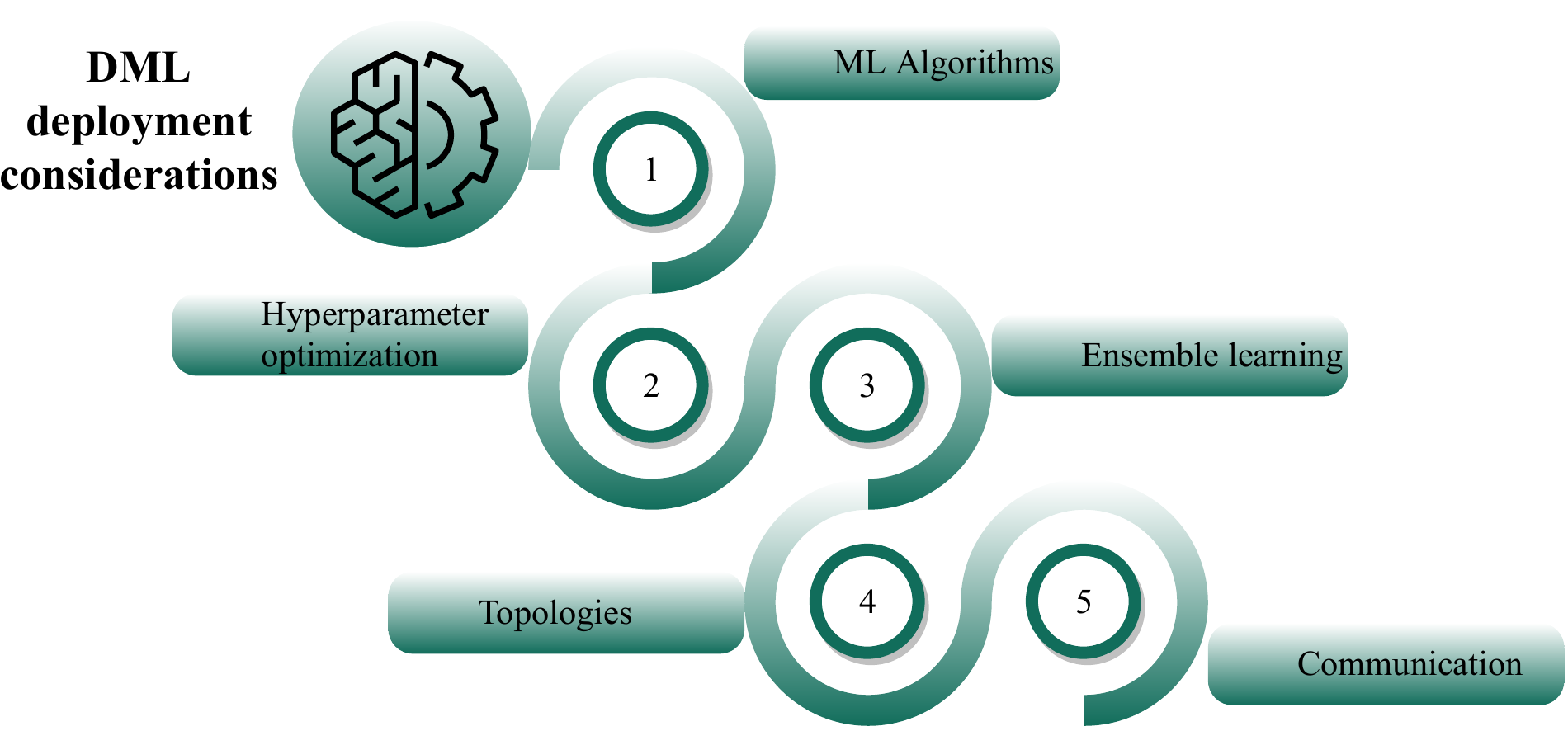}
    \caption{Roadmap of DML deployment. The following considerations are taken into account: (1) represents the set of ML algorithms, (2) denotes the problem of the choice of hyperparameters in the optimization process, (3) is an essential concept in DML deployment, which serves to combine several models to improve predictive performances, which shows the idea of ensemble learning, (4) represents the network typology considered in DML deployment namely centralized and decentralized, and (5) expresses the type of communication technique such as synchronous and asynchronous schemes.} 
    \label{fig:RoadMapDMLdeploymentCons}
\end{figure}

Multiple data instances, high dimensional data, model and algorithm complexity are among the reasons to scale ML \cite{bekkerman2011scaling}. When implementing DML, it is necessary to look to High-performance computing-style (HPC) hardware for the best performance, such as NVLink, InfiniBand networking, and GPUs with capabilities such as GPUDirect (more details in TABLE~\ref{tab:hardwareCons}).
\begin{table}[t]
   \caption{Brief description of some HPC-style hardware that must be considered while deploying DML.}
  \label{tab:hardwareCons}
  \begin{minipage}{\columnwidth}
  \small
  \begin{center}
 \begin{tabular}{p{3 cm} p{5 cm} }
\toprule
    HPC-style hardware&Brief description\\
    \midrule
      NVLink \cite{harris-2020}& High-speed GPU-to-GPU interconnect introduced by NVIDIA \\
      InfiniBand networking \cite{InfiniBand-2019}& HPC networking communication technology with extremely high performance and low latency \\
      GPUDirect \cite{GPUDirect-2021}& Magnum IO suite of technologies that improve data transportation and access for NVIDIA data center GPUs\\
\bottomrule
\end{tabular}
\end{center}
\bigskip
\end{minipage}
\end{table}
 Towards the deployment of DML, many considerations \cite{verbraeken2020survey} must be taken (see Fig.~\ref{fig:RoadMapDMLdeploymentCons}).

\textit{ML algorithm} is a construct that makes predictions from data and is based on three key ingredients: \textit{type}, \textit{goal}, and \textit{method}. The principal \textit{types} are represented by supervised learning \cite{bekkerman2011scaling}, unsupervised learning \cite{bekkerman2011scaling}, semi-supervised learning \cite{chapelle2009semi}, and reinforcement learning \cite{ben2019demystifying}. The \textit{goal} of machine learning is to find patterns in the data and make predictions based on those patterns, which are usually complicated. An optimal ML algorithm must have a \textit{method} that forces it to improve based on new input data in order to enhance its accuracy \cite{verbraeken2020survey}. These methods include SGD, which dates back to the 1950s \cite{robbins1951stochastic} and is an approximation of gradient descent (GD).

ML algorithm \textit{optimization} is divided into distinct families discussed in the literature \cite{beck2017first, bottou2018optimization, wright2015coordinate, brooks1998markov, ensor1997stochastic, bergstra2012random, shahriari2015taking}, each with their own set of methodologies. Our study focuses on first-order optimization algorithms, which date back to 1847 \cite{beck2017first}. They represent a set of ``methods that exploit information on values and gradients/subgradients\footnote{The subgradient method is an iterative method for tackling convex minimization problems and applied to a non-differentiable objective function. The observable differences from the standard gradient method are in step lengths which are fixed in advance, and the function value might grow in the subgradient method. Unlike the gradient approach, this is a descent method that uses exact or approximate line search \cite{boyd2003subgradient}.} of the functions comprising the model under consideration'' \cite{beck2017first}. The choice of algorithm \textit{hyperparameters} has a considerable influence on the performance of ML algorithms. Examples of hyperparameters are the learning rate, and momentum in SGD \cite{maclaurin2015gradient}. However, the optimal choice of hyperparameters depends on the problem domain, ML model, and dataset \cite{verbraeken2020survey}.

Emulating human nature on a combination of different viewpoints to make a critical decision is the origin of \textit{ensemble learning} \cite{sagi2018ensemble}, a  concept necessary for DML deployment. The predictive performance of a single model may not be as accurate as that of combining multiple models \cite{sagi2018ensemble, verbraeken2020survey}. Ensemble learning could be achieved by several methods like AdaBoost \cite{freund1997decision}, which focuses on leveraging data misclassified by previous models to train new models.

\textit{Network topology} is another important concept for designing a DML deployment, and the authors of \cite{verbraeken2020survey} presented four topologies: Centralized (Ensembling), Decentralized (Tree, Parameter Server), and Fully Distributed (Peer to Peer). In the next section, we discuss the segmentation of the topologies approved in our survey.

Despite the advantages of distributed systems in ML, such as using many processors to train a single model, which helps to reduce the training time, \textit{communication} is crucial in DML. The cost of communication between processors can affect the scalability of the system \cite{shi2018performance}, requiring data communication optimization. 
In a distributed system, identifying which tasks can be run in parallel, the order of task execution, and balancing the load distribution among the available machines must all be considered \cite{xing2016strategies}. Then, Bulk Synchronous Parallel \cite{xing2016strategies}, Approximate Synchronous Parallel \cite{hsieh2017gaia}, and other techniques may improve communication efficiency in terms of information exchange, which can ensure (fast/correct) model convergence with (faster/fresher) updates \cite{verbraeken2020survey}. In contrast, our survey focuses on synchronous/asynchronous algorithms, DML's most prevalent and commonly utilized methods.

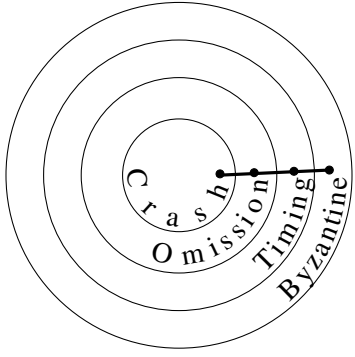
\begin{figure}[H]
  \centering
  \begin{tikzpicture}[>=stealth',join=bevel,auto,on grid,decoration={markings,
    mark=at position .5 with \arrow{>}}]
 
    \coordinate (structuralNode) at (2:2cm);
    \coordinate (originNode) at (0:0cm);

    \draw[-, very thick,black] (structuralNode.south) -- (0.5,0)
    node[pos=0]{ } 
    node[pos=0.1]{} 
    node[pos=0.4]{} 
    node[pos=0.7]{} 
    node[pos=1]{};

    \draw[fill,black] (barycentric cs:structuralNode=0.9,originNode=0.0) circle (1.5pt);
    \draw[fill,black] (barycentric cs:structuralNode=0.65,originNode=0.2) circle (1.5pt);
    \draw[fill,black] (barycentric cs:structuralNode=0.35,originNode=0.35) circle (1.5pt);
    \draw[fill,black] (barycentric cs:structuralNode=1.5,originNode=4.0) circle (1.5pt);

    \draw[black] (0,0) circle (2.3cm);
    \draw[black,postaction={decorate,decoration={raise=-2.5ex,text along path,text align=right,text={Byzantine}}}] (0,0) circle (1.8cm);
    \draw[black,postaction={decorate,decoration={raise=-2.5ex,text along path,text align=right,text={Timing}}}] (0,0) circle (1.3cm);
    \draw[black,postaction={decorate,decoration={raise=-2.5ex,text along path,text align=right,text={Omission}}}] (0,0) circle (0.75cm);
    \draw[white,postaction={decorate,decoration={raise=-2.5ex,text along path,text align=right,text={Crash}}}] (0,0) circle (0.25cm);
  
  \end{tikzpicture}
  \caption{Faults type in a distributed system. Crash: it causes loss of the internal state of the component. Omission: the component does not respond to some inputs. Timing: the component reacts too early or too late. Byzantine: an arbitrary behavior of the component.}

  \label{fig:FaultsTypes}
\end{figure}

Distributed systems face many challenges, including fault handling\cite{coulouris2005distributed}, which also applies to DML. Tolerating faults (hardware or software) is one of many techniques used to handle failures \cite{coulouris2005distributed}. Fault tolerance in distributed environments is defined as the ability of a system to continue performing its intended task in the presence of failures \cite{ahmed2013survey}. The types of possible failures\cite{jalote1994fault} in distributed systems are depicted in Fig.~\ref{fig:FaultsTypes}, and the Byzantine failure is the general one\cite{lamport1982byzantine}. When a component unexpectedly deviates from its intended behavior, it is referred to as a Byzantine failure \cite{mostefaoui2018randomized}. This undesirable conduct is known as adversarial, compromised, attack, malicious, and arbitrary failure. It can be malevolent or simply the consequence of a transient malfunction that changed the component's local state, causing it to behave in an unanticipated manner \cite{mostefaoui2018randomized}. Byzantine failure is described firstly in the synchronous distributed systems context \cite{lamport1982byzantine, pease1980reaching, raynal2010fault} before being examined in the asynchronous distributed systems situation \cite{attiya2004distributed, lynch1996distributed, raynal2012concurrent}.

\subsection{Problem formulation and motivation of the survey}
This study aims to analyze the proposed BFT approaches for DML, considering synchronous and asynchronous training processes in centralized and decentralized settings. Furthermore, examine to what extent BFT approaches in DML meet the requirements, such as Byzantine resilience methods, tolerance of Byzantine fault in the distributed SGD algorithm, preserving privacy in collaborative deep learning, etc., and the obstacles they confront.

The Byzantine faults in distributed systems include software, hardware, computation errors, and network problems as propagating wrong data between nodes, which apply to distributed machine learning, thus impacting negatively on the intended results. In more detail, the problem of Byzantine fault in DML can occur during training (concerns the adversarial gradients) as \cite{blanchard2017machine}, or inference (concerns the adversarial examples\footnote{Represent the maliciously generated inputs \cite{szegedy2014intriguing} and consists of small imperceptible changes to humans in a regular test sample, but they may misclassify the examples in many ML models, including deep neural networks \cite{yuan2019adversarial, wang2016theoretical, NEURIPS2020_00e26af6}}) like \cite{wang2016theoretical}. The former is our study on which we focus, where the sets of adversaries during training can know the information of other nodes and replace the data transmission among them with arbitrary values, which may break the robustness of distributed optimization methods used in DML. Several studies, including \cite{blanchard2017machine, weng2019deepchain, chen2018draco, el2018hidden, melis2019exploiting, hitaj2017deep}, have explained that an adversary in the training phase can inject erroneous or malicious data samples as false labels or inputs. For example, a Byzantine participant in the training process can select an arbitrary input, resulting in unwanted gradients. Hence,  as a possible consequence when aggregating, we may get incorrect gradients. Deducing sensitive and private information is another problem. In this context, adversaries might use the inference attack, for example, to extract sensitive model information \cite{nasr2019comprehensive}. Furthermore, they might change the model's parameters or alter the gradient in the incorrect direction, leading it to deviate arbitrarily, shifting the average vector away from the desired direction. These scenarios can disrupt the training process, where they serve to achieve unwanted gradients, which significantly impact the parameter update stage and, therefore, negatively impact the expected model outcome.

This study emphasizes Byzantine failure, which occurs when worker machines act arbitrarily, making data privacy more challenging to protect \cite{weng2019deepchain}. Higher accuracy, for example, is desired in several artificial neural network tasks, such as image recognition; this requires a huge number of data to train deep learning models, resulting in high computational costs \cite{gupta2016model, chilimbi2014project}. While distributed deep learning addresses the latter issue, privacy remains challenging throughout the training process. That is due to inferring critical training data using intermediate gradients when partitioned and stored separately \cite{song2017machine, melis2019exploiting, orekondy2018gradient}. In addition, escaping saddle points is a significantly more challenging task. As shown in \cite{yin2019defending}, the adversary may modify the landscape of the loss function around a saddle point. Another challenge is achieving optimal convergence that Byzantine attackers may prevent. Byzantine attackers can send distorted data or effects the algorithms to force them to converge to the attackers' choice-value \cite{cao2019distributed}. \textbf{Specifically, the studied problem at the core of our analysis is how to ensure the robustness of the SGD-based training algorithms in DML}.

\subsection{Scope of the survey}
On this hand, we define the scope of this survey as follows: \begin{itemize}
 \item Preliminaries of distributed machine learning 
\item Description of key techniques used in BFT for DML
\item Description of fundamental approaches proposed to deal with the BFT issue in DML.    
\end{itemize}
In subsequent sections, a discussion and a comparative analysis of the studied approaches are also given.

\subsection{Our contributions}

The purpose of this paper is to survey the proposed approaches targeted to tolerate Byzantine faults in the context of first-order optimization methods based on SGD. Although works on this subject have already been conducted, a complete study that includes the most current approaches, the classification of techniques used by these approaches, and the open problems in BFT in the DML is still lacking. The points that we will cover are provided in the following paragraph.

This study seeks to give a full overview of the BFT in the DML with preliminary definitions, achievements, and challenges conducive to providing the reader with a basic understanding of our proposed work. Therefore, the reader will be aware of current research trends in this scope and will be able to identify the most demanding concerns.

The description of the contributions to recent field literature in this work can be summarized as follows:
\begin{itemize}
\item Providing an overview with a classification of the techniques and approaches used to achieve solutions of BFT in DML
\item Identifying and examining solutions based on filtering schemes, coding schemes, and Blockchain technology to deal with BFT in DML with software environments
\item Providing insights into existing BFT approaches in DML and future directions   
\end{itemize}

\subsection{Survey organization}

The remainder of this paper is structured as follows: Section II provides a review of existing related works. Section III presents the essential terminologies related to DML. Section IV describes fault tolerance in DML frameworks and provides a detailed study of tolerant systems, focusing on Byzantine ones in SGD. Section V discusses and compares the set of research works presented in section IV. Finally, section VI offers the conclusion of our survey.
%============================================================
%%%%--------------------- Section 2---------------------------------
\section{Related Works}
In the literature, and to the best of our knowledge, despite extensive research in this field, we observe that only one overview has been published \cite{yang2020adversary}. Since there is no consensus on the usage of distributed and decentralized words in literature, Yang et al.\cite{yang2020adversary} chose to use them separately. They focused on the latest advances in Byzantine-resilient inference and learning. In this context, they provided the latest findings on the resilience of distributed/decentralized detection, estimation, and ML training algorithms in the presence of Byzantine attacks. The authors presented some Byzantine-resilient distributed stochastic gradient descent and decentralized learning approaches with numerical comparisons. Similarities and dissimilarities between distributed SGD approaches and those decentralized methods are provided in their work. However, this overview does not include any BFT approaches based on Blockchain or coding schemes in distributed machine learning to strengthen the discussion of the proposed BFT solutions. Moreover, the authors have not classified the techniques used by BFT approaches in the overview, in addition to very restricted approaches that have been explained.

We push the study further, making it more exhaustive and covering the full structure of BFT in DML, which might help to grasp its whole structure.

\section{Towards distributed machine learning}
%****************Subsection***********************
\subsection{Machine learning and Deep learning}

ML is one of the most successful Artificial Intelligence (AI) subfields, using algorithms capable of solving problems with progressively less human intervention. In other words, ML algorithms use data samples as inputs and make predictions as output values without any explicit programming of handcrafted heuristics.

\begin{figure}[h]
    \centering
   \includegraphics[width=110mm,scale=1]{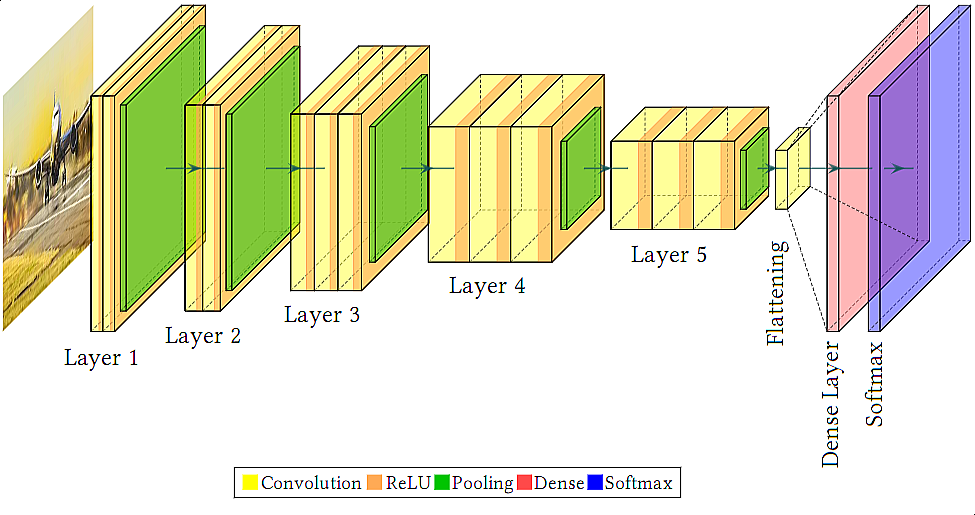}
       \caption{Deep Neural Networks Architecture adapted from \cite{haris-iqbal-2018}. The DNN architecture example of 8-layer (could be more or less). Layers 1 through 5 comprise a block of convolution operations, Rectified Linear Units (RELU) as an activation function, and pooling (downsampling) layers. Layer 6 represents the flattening layer. Layers 7 and 8 depict the dense layer and softmax layer, respectively.}
   \label{fig:DNN_Arch}

\end{figure}

One of the most successful ML subfields is Deep Learning (DL), which is based on Artificial Neural Networks (ANNs) inspired by the nervous systems. Deep Neural networks (DNN), which are ANNs with a very large number of layers, are the core of DL and are illustrated (see Fig.~\ref{fig:DNN_Arch}).
DL methods are ML methods based on artificial neural networks.

\begin{figure}[H]
    \centering
    \includegraphics[width=110mm,scale=0.8]{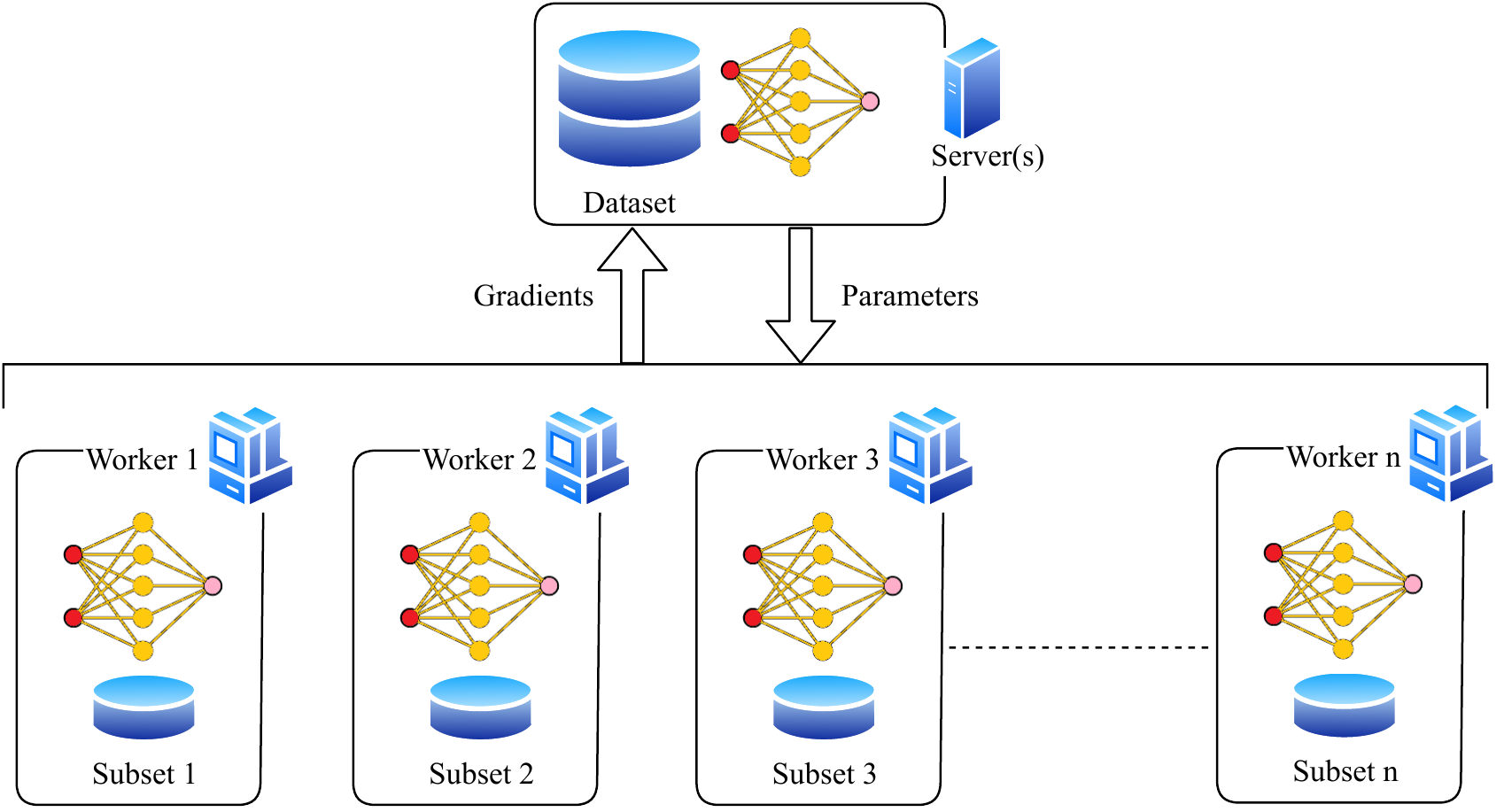}
   \caption{Schema of general distributed training. Distributed training is achieved with data-parallel and/or model-parallel techniques. The model is replicated across several (here $n$) workers in the former. Each worker performs its computations and then shares the parameter updates with the other workers. In the latter, the model is divided across $n$ workers, with each worker having a part of the model. Both techniques are based on synchronous or asynchronous updates, and more details are provided in the following section.}
    \label{fig:General_DistTrainArch}
\end{figure}

In large-scale learning, the complexity of DL models is growing \cite{srivastava2015training}, which has driven learning schemes that require an important set of computational resources. Consequently, distributing the computations among several workers in ML becomes necessary. A distributed training schema is illustrated in Fig.~\ref{fig:General_DistTrainArch}.

%%****************Subsection********************
\subsection{Distributed machine learning}

Distributed systems enable networked computers to cooperate in order to tackle complicated tasks requiring several computers, reducing the time it takes to achieve a good solution. In the case of ML, two standard techniques are used to divide the task among various computers (worker nodes) for efficient training: $ Data-parallelism $ or $Model-parallelism$. Both techniques are described in \cite{verbraeken2020survey} as follows: 

In $ Data-parallelism $, the data is partitioned on various worker nodes with a full model copy. All worker nodes apply the same algorithm to distinct data sets, with the results from each being a single coherent output somehow merged. 

In $Model-parallelism$, various worker nodes are in charge of processing accurate copies of the whole data sets on different parts of a single model. As a result, the model is comprised of all model parts.

The approaches mentioned above may also be used in conjunction \cite{xing2016strategies}. In the same context, several DML approaches have been developed as those based on the distributed SGD optimization algorithm \cite{agarwal2018cpsgd, basu2019qsparse, shi2019distributed}.

Optimization methods, topologies, and communication are among the main characteristics considered when deploying a DML, which we present in subsequent sections.

%********************Subsubsection***********************
\subsubsection{Optimization methods}
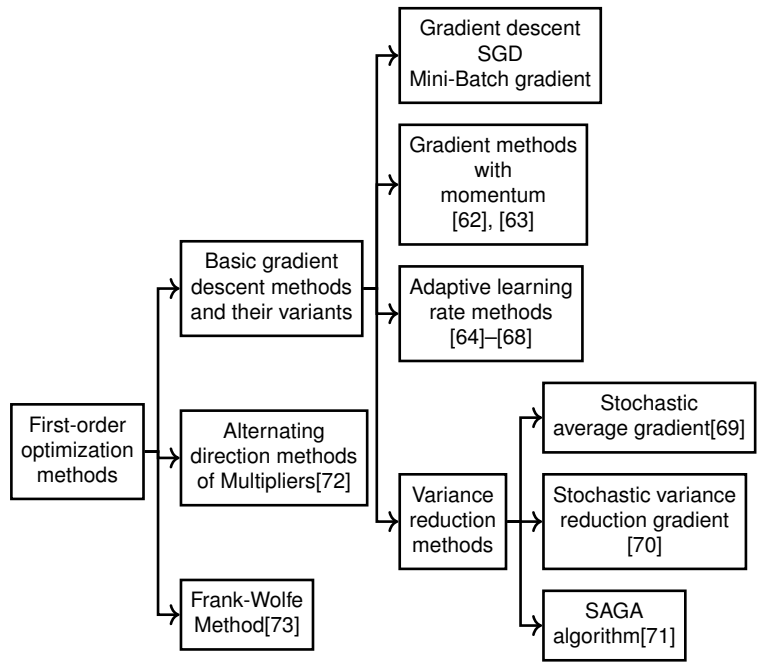
\begin{figure}

\begin{forest}
  for tree={
    font=\sffamily\scriptsize,
    line width=1pt,
    draw,
    anchor=west,
    child anchor=west,
    parent anchor=east,
    grow'=east,
    align=center,
    edge path={
      \noexpand\path[line width=1pt, ->, \forestoption{edge}]
      (!u.parent anchor) -| +(5pt,0) -- +(5pt,0) |- (.child anchor) \forestoption{edge label};
    },
  },
  s sep+=20pt,
  [First-order\\optimization\\methods
    [Basic gradient\\descent methods\\and their variants
          [Gradient descent\\SGD\\Mini-Batch gradient
          ]
          [Gradient methods\\with\\momentum\\\cite{polyak1964some, qian1999momentum}
          ]
          [Adaptive learning\\rate methods\\ \cite{duchi2011adaptive, zeiler2012adadelta, tieleman2012lecture, kingma2014adam, dozat2016incorporating}
          ]
           [Variance\\reduction\\methods
                [Stochastic\\average gradient\cite{le2012stochastic}]
                [Stochastic variance\\reduction gradient\\\cite{johnson2013accelerating}]
                [SAGA\\algorithm\cite{defazio2014saga}]
            ]
    ]
    [Alternating\\direction methods\\of Multipliers\cite{gabay1976dual}
    ]
     [Frank-Wolfe\\Method\cite{frank1956algorithm}
    ]
  ]
\end{forest}
    \caption{Types of First Order Optimization Methods in ML.}
    \label{fig:first_order_Opt_MLMeth}
\end{figure}

Optimization has drawn lots of interest and is considered a key component of ML. With the drastic rise of data and model complexity, optimization approaches in ML confront increasing hurdles. The commonly used optimization techniques are addressed in the literature \cite{bottou2018optimization,sun2019survey}, while we emphasis on the popular optimization methods family represented by first-order optimization methods (see Fig.~\ref{fig:first_order_Opt_MLMeth}).
ML optimization is the process of fitting out the hyperparameters (or evaluating the weights) using one optimization technique. In the literature, gradient descent is the most widely used optimization method \cite{banavlikar2018crop}. To reduce the prediction error of an ML model, GD can be used. GD minimizes the objective function $J(\theta)$ as follows \cite{ruder2016overview}:

\begin{algorithm}
\caption{Gradient Descent. \label{alg:alg1}}
\begin{algorithmic}
\STATE 
\REPEAT
\STATE \( \theta := \theta - \alpha \cdot \nabla_\theta J(\theta)\)
\UNTIL{convergence}
\end{algorithmic}
\end{algorithm}
where
\begin{itemize}
\item$\theta$: model parameters
\item$\alpha$: the learning rate that determines the steps size taken to achieve good local minima at the convergence.
\item$J$: objective function
\item$\nabla_{\theta} J(\theta)$: first-order gradient w.r.t. $\theta$.
\end{itemize}

\begin{figure}[H]
    \centering
    \includegraphics[width=150mm,scale=0.8]{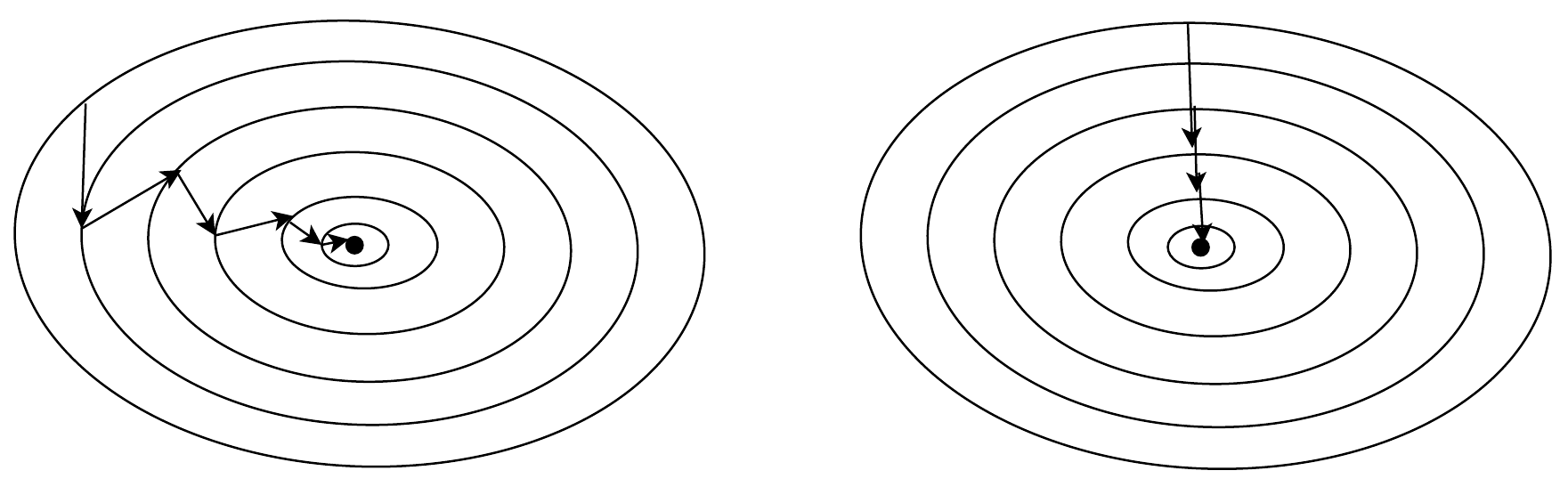}
    \caption{SGD (left) and GD (right) comparison adapted from \cite{shalev2014understanding}. In both GD and SGD, we iteratively update a set of parameters to minimize a loss function. GD traverses the full dataset once before each update, while SGD selects just one data point randomly. SGD frequently converges considerably quicker compared to GD. For large datasets, GD must recompute gradients for similar examples with each parameter update, while SGD eliminates this redundancy by updating one parameter at a time \cite{ruder2016overview}. Normally in terms of the number of steps, GD converges faster as it is a global estimate of a loss function. People also often suggest using larger batch in SGD when possible because it would be closer to GD.}
      \label{fig:SGDandGD}
  \end{figure}

Our interest is the SGD optimization method, which behaves according to the following algorithm and differs from the GD, as shown in Fig.~\ref{fig:SGDandGD}:

\begin{algorithm}
\caption{Stochastic Gradient Descent.}
\label{alg:alg2}
\begin{algorithmic}
\STATE 
\STATE Randomly shuffle training examples
\STATE for  \( i:=1, \ldots, m\{ \)
\REPEAT
\STATE \( \theta := \theta - \alpha \cdot \nabla J(\theta; x(i);y(i))\)\\
\STATE \( \} \)
\UNTIL{convergence}
\end{algorithmic}
\label{alg2}
\end{algorithm}
Where:
\begin{itemize}
\item  $m$: the number of training examples,
\item $x(i) ,y(i)$: the training examples.
\end{itemize}

The comparison discussed in \cite{ruder2016overview} and illustrated in Fig.~\ref{fig:SGDandGD} show that GD optimization can be costly in bulky datasets; GD computes the gradient of the cost function based on the complete training set for every update. Therefore, GD can be very slow and intractable in practice. As opposed to SGD, which uses one data point randomly to update parameters for each update. In addition, the convergence of GD is uncertain when there are several local minima (non-convex cases), which means when a global minima is required, the local minima may confine the training method to an undesired solution \cite{cetin1993global}. On the other hand, SGD avoids this problem by starting with a big step size and then reducing it after getting far enough away from the starting point \cite{kleinberg2018alternative}.

Despite the advantages of SGD, there are some challenges as to improve the SGD robustness \cite{dean2012large}, saddle points and selecting the learning rate \cite{ketkar2017stochastic} which have led to propose other variants of SGD like Momentum \cite{qian1999momentum}, Nesterov accelerated gradient \cite{nesterov1983method}, Adagrad \cite{duchi2011adaptive}, Adadelta \cite{ zeiler2012adadelta}, RMSprop \cite{tieleman2012lecture}, Adam and AdaMax \cite{kingma2014adam}, Nadam  \cite{dozat2016incorporating}. Additionally, for large-batch training, AdaScale SGD, another proposed variant that reliably adapts learning rates \cite{pmlr-v119-johnson20a}. Furthermore, in over-parameterized learning, an accelerated convergence rate was reached by MaSS \cite{liu2019accelerating} over SGD.

Naturally, SGD is a sequential algorithm \cite{bottou1998online}. When the crucial growth in training data size over the years (see Fig.~\ref{fig:GrowthData}) becomes a fact issue; hence, parallelized and distributed implementation is necessary for SGD and its variants. Distributed SGD implementations \cite{zhang2015deep} typically take the following form:
\begin{itemize}
\item The parameter server performs learning rounds to the training data \cite{li2013parameter}
\item Then, it diffuses the vector of parameters to the workers
\item Each worker calculates an estimate of the gradient
\item The parameter server aggregates the workers' calculation results
\item Finally, the parameter server updates the parameters vector
\end{itemize}

We cover in the next subsection the most common ways of communication used with SGD.

%********************Subsubsection********************
\subsubsection{Communication}
Synchronous / asynchronous approaches are the most used communication methods in the training process with distributed SGD.
\begin{figure}[h]
    \centering
    \includegraphics[width=140mm,scale=1.0]{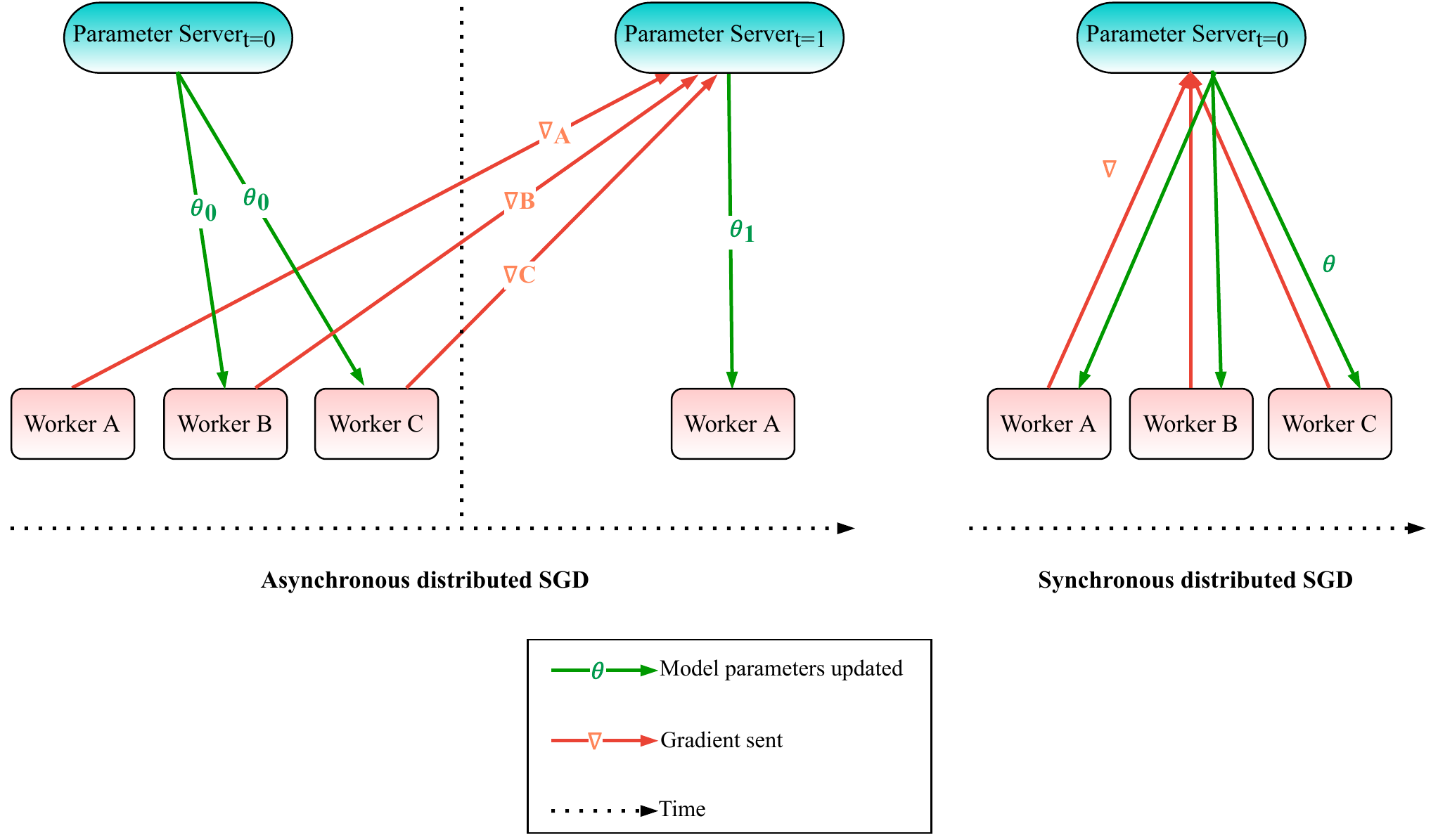}
    \caption{Asynchronous distributed SGD vs. Synchronous distributed SGD. Using three worker machines and a parameter server are required (the number of worker machines shown in the figure is an example; there may be more, as well as the parameter server, which is based on the topology used in the training process, as indicated in the next subsection). The computation happens across all workers and the parameter server, with workers submitting gradients to the parameter server, which delivers the updated model to the workers. The gradients transfer and updated model are occurred synchronously in synchronous distributed SGD (right) and asynchronously in asynchronous distributed SGD (left).}
    \label{fig:SynchAsynchDistSGD}
\end{figure}

Fig.~\ref{fig:SynchAsynchDistSGD} shows the synchronous distributed SGD vs. the asynchronous distributed SGD. The distinction between these methods is that no model update occurs until all workers have successfully calculated and delivered their gradients in the synchronous training case. In contrast to asynchronous training, no worker waits for a model update from another worker.

SGD optimization method (see Algorithm 2) works by determining the best weight parameters ($ \theta $) of the model by minimizing the objective function ($ J $). Thus, at each iteration, we minimize the error regarding the model's current parameters by following the gradient direction of the objective function of a single sample. 

When training deep models on large datasets, we need to distribute our computations using multiple machines. As we mentioned previously, synchronous and asynchronous methods are the most popular communication approaches used in the DML. We will discuss them with the distributed SGD.

Depending on the architecture, multiple workers process data in parallel. Each worker connects with the parameter server and processes a mini-batch of data dependently on the others in the synchronous approach and independently of the others in the asynchronous approach \cite{chen2016revisiting}.

\paragraph{\bf{ Synchronous distributed SGD }} 
Several works \cite{das2016distributed, zhang2018adaptive, zhao2019dynamic} used the synchronous method. The computation in the synchronous approach is deterministic and straightforward to implement. In every iteration, each worker calculates the model gradients in parallel using various mini-batches of data. Then, the parameter server waits for all workers to communicate their gradients before aggregating them and transmitting the updated parameters to all workers \cite{chen2016revisiting}.

\paragraph{\bf{ Asynchronous distributed SGD }}
The works \cite{lian2018asynchronous, reddi2015variance, zhang2013asynchronous} among the research papers are based on the asynchronous method. Furthermore, the asynchronous communication steps are presented in \cite{chen2016revisiting}. First, the worker obtains from the parameter server the most recent model parameters required to process the current mini-batch. It then computes gradients of the loss concerning these parameters; and finally, these gradients are returned to the parameter server, which updates the model accordingly.

In reality, when a worker retrieves the model's parameters while other workers update the parameter server, an overlap may occur, resulting in inconsistent results. Furthermore, there is no guarantee that model updates may have happened while a worker calculates its stochastic gradient; consequently, the resulting gradients are stale and often calculated regarding out-of-date parameters \cite{chen2016revisiting}.\\

The two most common distributed machine learning methods (synchronous/asynchronous) are built within a (centralized/decentralized) setting that we develop in the subsequent section. 

%***************Subsubsection*****************************
\subsubsection{Topologies}

We mentioned in the introduction four possible topologies discussed in \cite{verbraeken2020survey} based on the degree of distribution\footnote{The degree of distribution is the frequency distribution of a network system's node degree over the whole communication network.}. Due to the lack of agreement on the definitions of distributed and decentralized  \cite{yang2020adversary}, we centered our survey on most topologies used in BFT for DML. We classified them into centralized and decentralized (multiple parameter servers, fully distributed) settings.
\begin{figure*}[!t]
\centering
\subfloat[]{\includegraphics[width=1.65in]{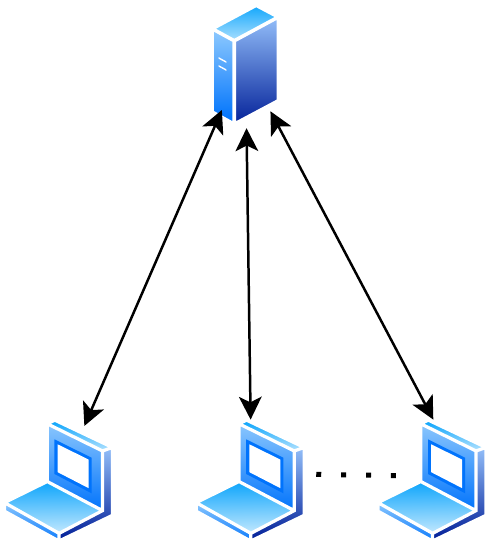}%
\label{fig_first_case}}
\hfil
\subfloat[]{\includegraphics[width=2.4in]{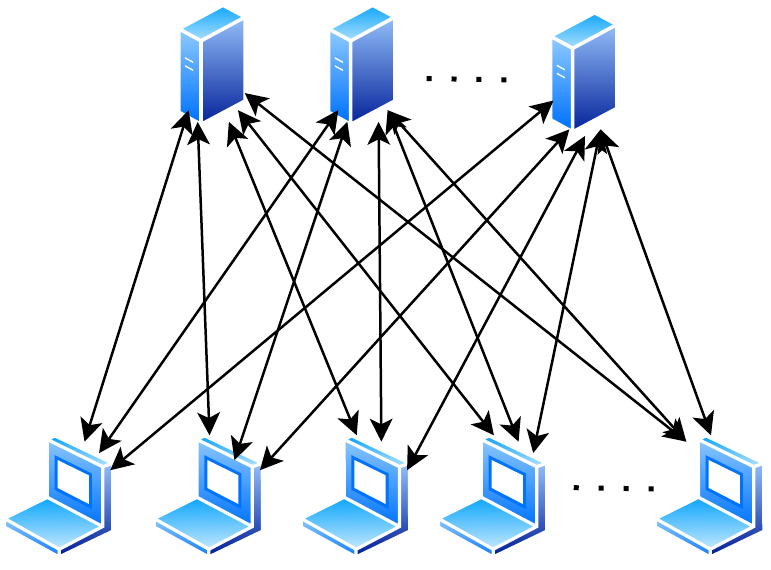}%
\label{fig_second_case}}
\subfloat[]{\includegraphics[width=1.9in]{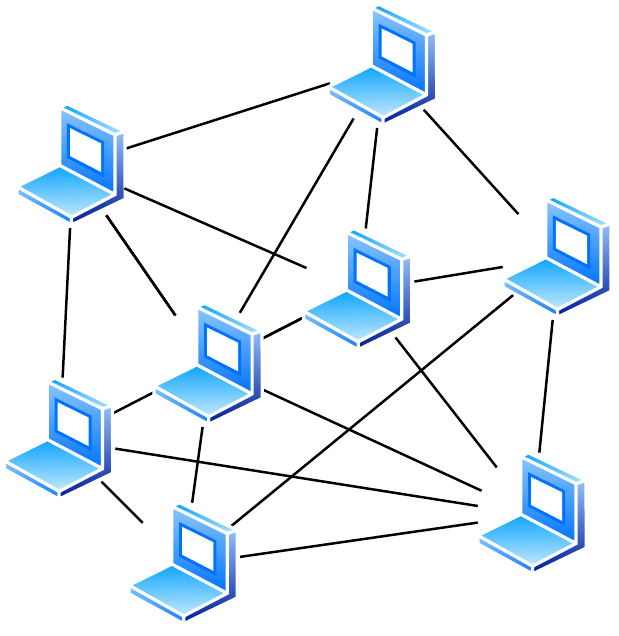}%
\label{fig_thi_case}}\\
\begin{tabular}{|cl|}
\hline
\includegraphics[width=8pt]{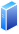} & \large Server \\
\includegraphics[width=15pt]{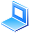} & \large Worker \\
\hline
\end{tabular}
\caption{(a) Centralized setting, which shows a central node computing (server) and worker nodes connecting with the server (b) Decentralized setting, the case of multiple parameter servers that shows several severs connecting with a decentralized set of workers (c) Decentralized setting, case of fully distributed setting that shows independent nodes that collaborate the solution together.}
\label{fig:settings}
\end{figure*}
\begin{enumerate}
\item \verb| Centralized setting|: the centralized setting is the classical one of the distributed machine learning paradigm, consisting of the parameter server model \cite{li2014scaling}, where there is a central node computing (server) and a set of data holders (worker nodes) that connect to the server (see Fig.~\ref{fig:settings}.a). The worker nodes perform the gradient calculation locally and stream it to the server that updates the model parameter.

\item \verb| Decentralized setting|: to eliminate single points of failure in a centralized setting, there is a need to use a decentralized setting that excludes the necessary central entity (server) in the centralized setting. Therefore, the decentralized setting focused on the cooperation among entities to produce the final result \cite{yang2020adversary}.

\paragraph{\bf{ Decentralized setting based on multiple parameter servers }}
in this setting, instead of using one parameter server to manage parts of the model, multiple centralized parameter servers are used with a decentralized set of workers \cite{verbraeken2020survey} (see Fig.~\ref{fig:settings}.b).

\paragraph{\bf{ Fully distributed setting }}
there are no central nodes in a fully distributed setting, so there are no single points of failure. This setting can be represented by a directed or undirected graph, where the workers communicate with each other and transit their parameter vectors to their neighboring nodes. Once the parameter vectors have been received, each worker node calculates the gradient locally and broadcasts it to its neighbors. Finally, the worker nodes aggregate the training model parameter from each neighboring worker node and update the model \cite{guo2020towards, lian2017can}. A \textit{peer-to-peer} network is an example of implementation of this setting (see Fig.~\ref{fig:settings}.c) also the based topology of \textit{Blockchain} technology, the core technology used to create the cryptocurrency \cite{nakamoto2008peer}. In other terms, Blockchain is a peer-to-peer distributed ledger technology that enables secure transactions and a bond of trust with its participating users \cite{dave2019survey}. Some of the works given in this survey rely on this technology in their proposed approaches, such as \cite{weng2019deepchain, chen2018machine, lugan2019secure, zhao2019mobile, rathore2019blockdeepnet}.

\end{enumerate}

%================================================================
%---------------------------Section 3----------------------
\section{Byzantine fault tolerance in distributed machine learning}

\subsection{Fault tolerance in distributed machine learning}

A system is fault-tolerant if it can detect and eliminate fault-caused errors or crashes and automatically recover its execution. Following the growth of data and the necessity of an efficient system in many fields, including machine learning, fault tolerance is becoming increasingly crucial \cite{zhou2003evolving, leung2008fault, simon1995fault,arad1997fault}.
The checkpoint/restart is the simplest form of fault tolerance in machine learning \cite{ben2019demystifying}, and the Self-Correcting Algorithm Recovery (SCAR) framework was developed by Aurick et al. \cite{qiao2019fault} among approaches that used this form. The presented framework is based on the parameter server architecture (PS)\cite{li2013parameter, ho2013more,li2014communication}. It minimizes the size of the perturbation caused by failure, which is reduced by 78\% to 95\% \cite{qiao2019fault} compared to fault tolerance training algorithms and models with a traditional checkpoint.
The increased number of proposed Byzantine fault-tolerant techniques is proportional to the number of different software faults that rise in tandem with various malicious attacks. 
In this paper, we provide a comparative study among the works in the literature that aim to tolerate Byzantine fault in the first-order optimization methods, especially in the distributed SGD optimization methods.

\subsection{Byzantine fault tolerance in first-order optimization methods}
We classified the Byzantine fault tolerance approaches that deal with first-order optimization methods, precisely the SGD method, into two categories: synchronous and asynchronous training approaches. Based on these two categories shown in Fig.~\ref{fig:SynchAsynchDistSGD}, we present a set of proposed algorithms, which are analyzed as follows:

%----------------------------------------------------------------------------
\subsubsection{Synchronous training approaches VS Asynchronous training approaches}

Synchronous stochastic gradient descent is used in the training process when worker communication is dependent on the parameter server. In this case, the algorithms must wait for the parameter server to aggregate all the gradients sent by the workers before updating the new model and moving on to the next iteration. However, asynchronous stochastic gradient descent is used when worker communication with the parameter server is independent. In this situation, the algorithm does not wait until the collection of worker gradients is complete so that the settings server can update the new model and move on to the next iteration.\\

\begin{figure}[h]
    \centering
    \includegraphics[width=110mm,scale=1.0]{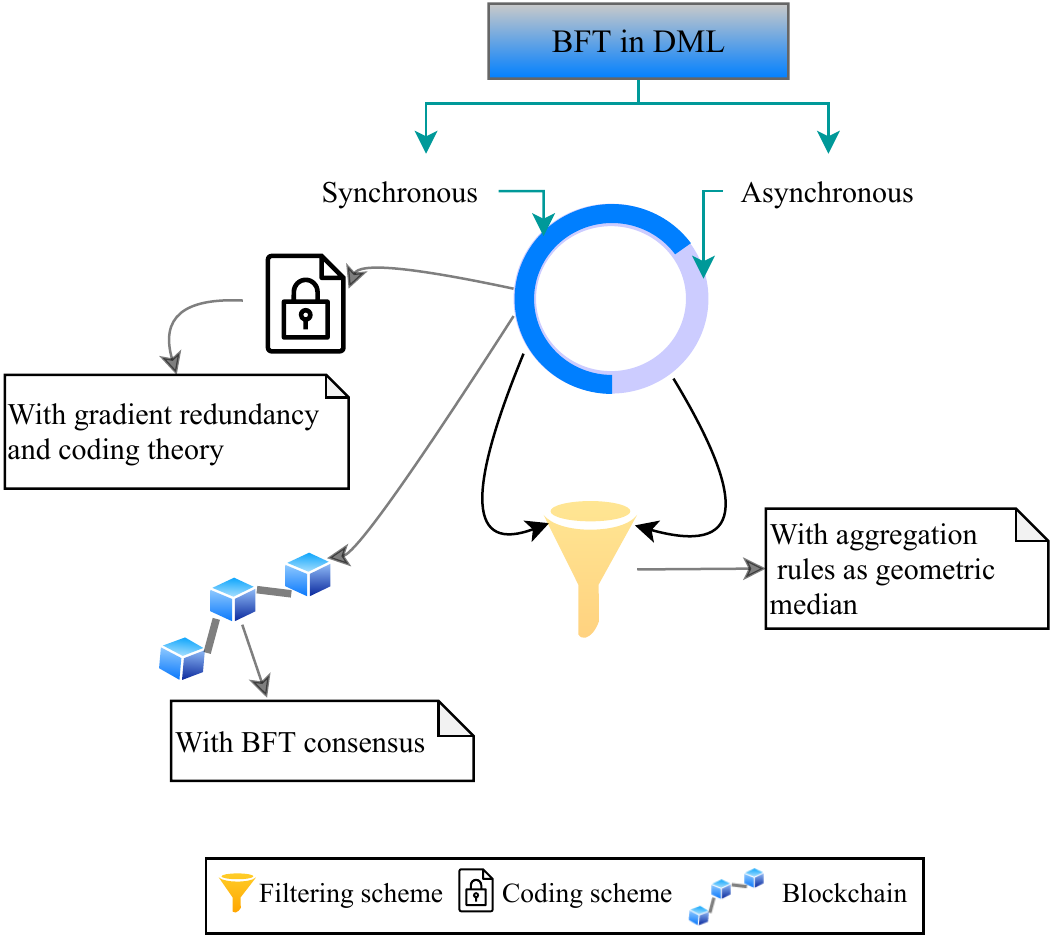}
    \caption{Brief description of the different types of techniques used in BFT in DML. Compared to the asynchronous training method, a large part of BFT approaches in DML use the synchronous training method. Synchronous approaches use various techniques, while asynchronous approaches use the filter scheme technique.}
    \label{fig:BFT_typesTechniques_descr}
\end{figure}

We classify the techniques used in first-order optimization methods for synchronous/asynchronous Byzantine fault tolerance into three types, as shown in Fig.~\ref{fig:BFT_typesTechniques_descr}.

\begin{enumerate}
\item \textit{\textbf{Byzantine fault tolerance based on filtering schemes}}: the aggregation rule as a geometric median \cite{cohen2016geometric} is used to calculate the average of the input vectors \cite{el2020robust} in the learning process. In the distributed training process and before the model update phase, aggregation must be passed by collecting the gradients to compute their average. Most of the BFT approaches in the literature are based on the filtering schemes presented by gradient filters (aggregation rules). In this type of method, the mistrustful gradients are filtered in some approaches before averaging by existing robust aggregation rules \cite{alistarh2018byzantine, xie2018zeno} as those based on median or mean \cite {tianxiang2019aggregation}. Another way to be accredited by other works is to use a robust aggregation rule that replaces the averaging aggregation operation \cite{blanchard2017byzantine, blanchard2017machine}. Other schemes are used in several ways, as \cite{su2018securing}, which use the iterative filtering based on robust mean estimation \cite{steinhardt2018resilience}, and the norm filtering used by \cite{gupta2019byzantine_b, gupta2019byzantine_a, li2019rsa} in addition to General Update Function (GUF) used with average aggregation by \cite{guo2020towards}.

\item \textit{\textbf{Byzantine fault tolerance based on coding schemes}}: these schemes are also known as redundancy schemes in some literary works. The two names come from the combination of redundant gradient calculations and the encoding scheme. The basic idea of the redundant mechanism is that each node evaluates a redundant gradient instead of a single gradient. More precisely, each computation node has a redundancy rate, representing the average number of assigned gradients. So, in the presence of Byzantine workers, the parameter server can always recover the sum of the correct encoded gradients per iteration \cite{chen2018draco}. The coded gradient presents the phase of coding theory, as shown in \cite{zaharia2008improving}. The coding-theoretic techniques have been used before but not in the case of Byzantine fault tolerance; instead, they have been used to speed up distributed machine learning systems\cite{lee2017speeding}, gradient coding \cite{tandon2017gradient, raviv2018gradient}, encoding computation \cite{dutta2016short}, and data encoding\cite{karakus2017straggler}. 

\item \textit{\textbf{Byzantine fault tolerance based on Blockchain}}: Blockchain was presented as a decentralized system characterized by auditability, privacy, and persistence \cite{zheng2017overview}. When a faulty component in a computer system exhibits an arbitrary behavior, we discuss the Byzantine faults. This issue introduces firstly by the Byzantine Generals Problem \cite{lamport1982byzantine}, an example of consensus methods to tolerate this type of problem. It is known as Byzantine fault tolerant consensus protocol. "Proof of Work" is the principal consensus protocol of Bitcoin's public Blockchain to tolerate Byzantine faults \cite{nakamoto2008bitcoin}. Blockchain technology is used in machine learning systems to overcome Byzantine challenges and preserve security and privacy.
\end{enumerate}

\subsection{BFT approaches in first-order optimization methods}
This section presents various BFT approaches proposed in the literature. We introduce the principle of each approach, the BFT method used, with some advantages and weaknesses. Sections \textit{1} and \textit{2} discuss BFT approaches in synchronous training. Section \textit{1} is centered on the SGD method, and approaches based on other ML optimization methods shown in section \textit{2}. The BFT approaches in asynchronous training are presented in section \textit{3}, while section \textit{4} introduces the BFT approaches in partially asynchronous training.\\

We use some abbreviations which are not standard. The authors chose these abbreviations according to the proposed approach.\\

We summarize the notations frequently used in the next subsections as follow:
\begin{itemize}
    \item $n$: The total number of worker machines 
    \item $f$: Byzantine workers
    \item $d$: The model dimension (represents the dimension of the parameter vector)
    \item $T$: The set of iterations in an iterative algorithm
\end{itemize}

\subsubsection{Synchronous training BFT approaches based initially on the SGD}

A brief description of each proposed approach is in the Table ~\ref{tab:CategorizationSynBFT}, along with the kind of technique it lies in.

\begin{table}[h]
  \caption{Categorization of synchronous training BFT approaches based on the SGD. Each BFT synchronous training approach is presented with its basic technique (filtering scheme, coding scheme, Blockchain) and a summary of its main idea. \textbf{Zoom the PDF for better visualization}.}
  
  \label{tab:CategorizationSynBFT}

\resizebox{1.0\columnwidth}{!}{
\centering
\begin{tabular}{|l|c|c|c|l|}
\hline
\cline{2-4}
    Approach& \multicolumn{3}{ c| }{Technique Type}&Brief Description\\
\cline{2-4} & Blockchain&Filtering scheme&Coding scheme& \\
    \hline
     DeepChain&$ \bullet $& & &A robust platform for secure collaborative deep training\\
     \hline
    Krum and Multi-Krum& &$ \bullet $& &An aggregation rule to guarantee the algorithm convergence despite Byzantine workers\\
         \hline
    ByzantineSGD& &$ \bullet $& &A variant of SGD to guarantee the algorithm convergence to an optimum convex objective solution\\
         \hline
    GBT-SGD& &$ \bullet $& &Three median-based aggregation rules tolerating Byzantine failures for distributed synchronous SGD\\
         \hline
    Tremean,Phocas& &$ \bullet $& &Two trimmed-mean-based aggregation rules to guarantee the dimensional Byzantine-resilient\\
         \hline
    DRACO& & &$ \bullet $&A framework for dealing with malicious computation nodes using algorithmic redundancy and coding theory concepts\\
         \hline
    Bulyan& &$ \bullet $& &A robust algorithm that enhances the Byzantine resilience of SGD\\
         \hline
    Zeno& &$ \bullet $& &A robust aggregation rule tolerating Byzantine failures for distributed synchronous SGD\\
         \hline
    LearningChain&$ \bullet $& & &A privacy-preserving and secure decentralized learning framework\\
         \hline
   RSA& &$ \bullet $& &An efficient variant of SGD for distributed learning from heterogeneous datasets under the Byzantine attacks\\
        \hline
   SIGNSGD& &$ \bullet $& &An algorithm for robust, communication-efficient learning, which based on the sign of workers gradient vector and majority vote aggregation\\
        \hline
   AGGREG-ATHOR& &$ \bullet $& &A framework that implements state-of-the-art robust distributed SGD\\
        \hline
   GuanYu& &$ \bullet $& &An algorithm that deals with Byzantine parameter servers and Byzantine workers\\
        \hline
   Trimmed mean& &$ \bullet $& &A mean-based aggregation rule that addresses the Byzantine problem and meets the Byzantine dimensional resilience criterion\\
        \hline
   FABA& &$ \bullet $& &A fast aggregation algorithm that eliminates the outliers in the uploaded gradients so that they are close to the true gradients\\
        \hline
   BRIDGE& &$ \bullet $& &A decentralized learning approach for learning effective models while addressing a specific number of Byzantine nodes\\
        \hline
   LICM-SGD& &$ \bullet $& &A Lipschitz-inspired coordinate-wise median SGD algorithm that tolerates Byzantine workers without knowing how many they are\\
        \hline
   Gradient-Filter CGC& &$ \bullet $& &A parallelized SGD method to deal with Byzantine workers in synchronous settings, based on comparative gradient clipping\\
        \hline
   LIUBEI& &$ \bullet $& &A Byzantine-resilient algorithm that filters out Byzantine servers while avoiding communication with all servers\\
        \hline
   DETOX& &$ \bullet $& $ \bullet $&A framework for Byzantine-resilience distributed training using algorithmic redundancy and a robust aggregator\\
        \hline
   RRR-BFT& & &$ \bullet $&Two coding schemes (deterministic and randomized) to tolerate Byzantine fault in the parallelized-SGD learning algorithm\\
        \hline
   Stochastic-Sign SGD& &$ \bullet $& &A framework based on SIGNSGD to deal with heterogeneous data distribution\\
   \hline
\end{tabular}

\bigskip
}
\end{table}

\begin{enumerate}
\item \textit{BFT approaches based on filtering schemes:}
\paragraph{\textbf{Krum and Multi-Krum}}
Blanchard et al. \cite{blanchard2017byzantine, blanchard2017machine} presented the problem of tolerating Byzantine failure in the distributed SGD algorithms. The authors identified that no aggregation rule in the previously proposed approach of \cite{blanchard2017byzantine} can tolerate one Byzantine failure. To tolerate $ f $ Byzantine workers, the authors defined a Byzantine resilience property for the parameter server's aggregation rule based on two conditions:
\begin{enumerate}
\item Lower bound on the scalar product of the vector F (the output vector) and g(the real gradient): the Euclidean distance generated by the aggregation rule between F and g must be minimum.
\item The moments of the correct gradient estimator G control the F's moments: that means this condition allows the aggregation rule to do this control instead of making it by the bounds on the moments of G that control the SGD dynamics' discrete nature's effects\cite{bottou1998online}.
\end{enumerate}
The defined resilience property is guaranteed by integrating the majority-based and squared-distance-based methods so that the authors may pick the vector closest to its $n - f $ neighbors, and the one whose squared distance is the smallest to its $n - f$ vectors; this is the underlying principle behind Krum, the proposed aggregation rule.
Krum satisfies the Byzantine resilience property and defends against  \textit{Gaussian} and \textit{Omniscient} Byzantine workers, as just one attack prevents SGD from becoming convergent. To increase the SGD algorithm's ability to deal with attacks from adversarial workers, Krum uses a convex function combined with a robust aggregate rule (geometric median), with the assumption that $2f + 2 < n$.

In the experimental evaluation, the authors proposed Multi-Krum, a variant of Krum. Multi-Krum calculates the score as in the Krum function for each vector. Then, in multiple rounds of Krum, it takes the average on several selected vectors. The time complexity achieved by Krum is $O(n^2d)$.\\

\textit{Weaknesses:} Krum's non-convex function did not converge, and the Byzantine worker must be less than $n/2$.

\paragraph{\textbf{ByzantineSGD}}
Alistarh et al. \cite{alistarh2018byzantine} presented the ByzantineSGD algorithm that solves the distributed stochastic convex optimization problem. Despite the $f$ malicious Byzantine workers, the authors want ByzantineSGD to converge to an optimum convex objective solution. To identify honest workers, ByzantineSGD aggregates workers' reports using the median of means, compares their shared gradients with the median, and then updates the parameter according to the correct gradient information.

ByzantineSGD optimally achieves two fundamental criteria, sample complexity and computational complexity. It provides high accuracy with few data samples (sample complexity) and preserves runtime speedups (computational complexity) by distributing computation. When the dimension grows, ByzantineSGD stays useful and converges. Thus, the authors' proposed approach proved to be highly robust.\\ 

\textit{Weaknesses:} ByzantineSGD was not applicable in the decentralized model, and it was less practical due to its demand for the estimated upper bounds of the stochastic gradients' variances.

\paragraph{\textbf{GBT-SGD}}
Xie et al. \cite{xie2018generalized} proposed a GBT-SGD (Generalized Byzantine-tolerant SGD), which displays three median-based aggregation rules, to tolerate the Byzantine failure in the distributed synchronous SGD. The attacks may corrupt the model training, where it converges slowly or to a bad solution. The proposed aggregation rules based on median (GeoMed: geometric median, MarMed: marginal median, and MeaMed: mean around the median) prove the convergence to the good quality solutions with $f < n/2$ for each dimension, and this property, namely "dimensional Byzantine resilience." 

The main advantages of this approach are ensuring aggregation rules convergence despite the Byzantine values and the low computation cost. According to the authors \cite{xie2018generalized} is the first approach that presents the property of dimensional Byzantine resilience (DBR) and generalized Byzantine failures (GBF) for synchronous SGD; also, it applies under non-convex settings.\\

\textit{Weaknesses:} GBT-SGD aggregation rules become incapable if more than half of the existing workers are Byzantine.

\paragraph{\textbf{Phocas, Trmean}}
Xie et al. \cite{xie2018phocas} presented the dimensional Byzantine-resilient algorithms (Phocas and Trmean). These Byzantine-resilient aggregation rules are based on the generalized of the classic Byzantine resilience property \cite{blanchard2017machine}. The first condition is that the number of correct workers is higher than Byzantine workers \cite{xie2018generalized}. The authors defined Trmean's trimmed-mean-based aggregation rule, which describes another aggregation rule, Phocas. The proposed rules use fewer assumptions and guarantee convergence in the synchronous SGD, which is proved for the general smooth loss functions. 

According to the results of the experiments, Phocas and Trmean are performing and attaining linear time complexities.\\

\textit{Weaknesses:} Trmean's worst is constantly under omniscient attack, which slows convergence and makes it difficult in practice.

\paragraph{\textbf{Bulyan}}
El Mahdi et al. \cite{el2018hidden} addressed the problem of SGD convergence to an ineffectual model. The authors discussed poisoning attacks in high dimension $d\gg 1$ with non-convex loss functions. In this context, they showed the deficiencies of existing approaches that have proven resilient despite the presence of Byzantine workers \cite{chen2017distributed, blanchard2017machine}. The Bulyan algorithm, a defensive hybrid mechanism, is proposed as a solution. To pick the true gradients, Bulyan uses the geometric-median-based method; to determine the update direction, it uses the coordinate-wise trimmed mean method. 

Bulyan reduces a Byzantine worker's leeway to an upper-bounded $O(1\sqrt{d})$ and converges to successful learning models comparable to the reasonable benchmark.\\ 

\textit{Weaknesses:} Bulyan does not treat all the attack problems.

\paragraph{\textbf{Zeno}}
Xie et al.\cite{xie2018zeno} generalized the works \cite{xie2018generalized, xie2018phocas} and proposed Zeno. Zeno is a robust aggregation rule that suspects the Byzantine workers in the SGD's distributed synchronous by using the weakest assumption (the existence of one correct worker, at least). The authors used the stochastic first-order oracle to determine a score for each candidate gradient estimator. The candidates were then sorted based on the loss function' estimated descent and the magnitudes of the updates. Zeno aggregates the candidates based on the most elevated scores.

When there is at least one non-Byzantine worker, the authors proved Zeno convergence, which is as quick as SGD. Increasing the number of non-Byzantine workers will decrease variance, as well as the time complexity (theoretically). The experimental results show that Zeno performs effectively under Sign-flipping and Omniscient attacks.\\

\textit{Weaknesses:} Zeno requires the presence of a validation dataset on the parameter server, which is impractical in some circumstances.

\paragraph{\textbf{RSA}}
Li et al. \cite{li2019rsa} proposed RSA (Byzantine-Robust Stochastic Aggregation methods) to mitigate the Byzantine attacks' negative effects and make the learning task robust. RSA is a collection of resilient stochastic algorithms created to prevent Byzantine attacks on heterogeneous data sets. It used an aggregation of models to obtain a consensus model. In RSA, no assumption that the data are independent and identically distributed ( i.i.d. ).
Firstly, and by using subgradient recursions, the authors proposed $\ell_1$-norm RSA (Byzantine-robust stochastic aggregation), which is robust to arbitrary attacks from Byzantine workers and reduces every workers' variable to be close to the master's variable. The authors then expanded the $\ell_1$-norm regularized problem to the $\ell_p$-norm regularized problem. The latter helps alleviate the negative impact of Byzantine workers.

The resultant SGD-based (RSA) methods have a near-optimal convergence rate, where the learning error depends on the number of Byzantine workers. Furthermore, in Byzantine-free conditions, both SGD and RSA have a comparable convergence rate, which is an optimum solution at a sub-linear convergence rate.\\

\textit{Weaknesses:} the numerical test shows that $\ell_\infty$-norm does not provide competitive performance.

\paragraph{\textbf{SIGNSGD}}
Bernstein et al. \cite{bernstein2018signsgd} explored the SIGNSGD algorithm to obtain robust learning. Instead of communicating their gradients, the set of workers uses the SIGNSGD algorithm to send only the sign of their gradient vector to the server. By majority vote, the global update is decided. The authors assumed that (fast algorithmic convergence, good generalization performance, communication efficiency, and robustness to network faults) goals that the SIGNSGD achieved as follows:
\begin{itemize}
\item The authors proposed compressing all communication among the server and the workers to one bit to attain communication efficiency. 

\item Methods based on folk signs are performing well, which means SIGNSGD can quickly achieve algorithmic convergence. SIGNSGD provides a theoretical basis for fast algorithmic convergence in the mini-batch. The authors indicated that its behavior changes theoretically from a high to a low signal-to-noise ratio.

\item SIGNSGD used majority vote theory to aggregate gradients to achieve robustness to network faults, implying that no one worker has too much power. Despite the assumed model that considers robustness to Byzantine workers or workers that reverse their gradient estimate, it is not the most general. SIGNSGD, on the other hand, may achieve BFT while providing instantaneous good generalization performance.
\end{itemize}

Empirically, the authors prove the convergence of SIGNSGD in large and mini-batches. SIGNSGD decreases communication by iteration compared to distributed SGD, and when more than half of the workers are Byzantine, SIGNSGD proves the robustness of majority voting, unlike SGD.\\

\textit{Weaknesses:} the majority voting may need further optimization to avoid a single machine becoming a communication bottleneck. Furthermore, SIGNSGD has a test set gap and fails to converge when the compute units are heterogeneous.

\paragraph{\textbf{AGGREGATHOR}}
Damaskinos et al. \cite{damaskinos2019aggregathor} presented AGGREGATHOR framework to leverage more workers to speed up learning. AGGREGATHOR was built around TensorFlow\footnote{TensorFlow is one of the deep learning frameworks. These deep learning frameworks may be interfaces or libraries that serve data scientists and ML developers to develop deep learning models. Other frameworks include PyTorch, Keras, MXNet, and more.} and aimed to achieve two goals:
\begin{itemize}
\item The robust and distributed training on TensorFlow will be faster in the development and test cases.
\item The authors want to enable the deployment of Byzantine-resilient learning algorithms exterior to the academic environment.
\end{itemize}
Each worker might communicate with the server duplications and use the model sent by 2/3 of them. Because gradient aggregation rules (GAR) and updating the model in the server are deterministic, good servers will provide similar models to workers.

The following are some of the proposed framework benefits :
\begin{itemize}
\item In the case of a congested network, AGGREGATHOR uses UDP instead of TCP communication protocol, which provides a further speed up and makes TensorFlow more efficient. That helps to avoid losing accuracy.
\item Through MULTI-KRUM \cite{blanchard2017byzantine}, AGGREGATHOR ensures the first level of robustness (weak resilience), and by BULYAN \cite{el2018hidden}, it provides the second level of robustness (strong resilience).
\item AGGREGATHOR addresses the TensorFlow weakness to achieve Byzantine resilience. 
\item AGGREGATHOR is used securely in the training process to distribute any ML model developed for TensorFlow.\\
\end{itemize}

\textit{Weaknesses:} the overhead of AGGREGATHOR. Furthermore, the interaction between gradient descent's specificity and AGGREGATHOR's state machine replication technique will be difficult to implement efficiently.

\paragraph{\textbf{GuanYu}}
El-Mhamdi et al. \cite{el2019sgd} presented GuanYu, which is theoretically the first algorithm that tolerates Byzantine parameter servers and Byzantine workers. For each step of synchronous distributed SGD and to update the parameter vector, GuanYu uses several gradient aggregation rules (GAR) to aggregate all worker gradient estimates into one gradient. The authors used a contraction argument, taking advantage of the geometric properties of the median in high dimensional spaces, to demonstrate the Byzantine resilience of the proposed algorithm. Within each non-Byzantine server, this argument aims to avoid any drift on the models (with probability 1).

GuanYu achieves optimal convergence and asynchrony. It was also built on TensorFlow and deployed in a distributed configuration. The authors demonstrated that GuanYu could tolerate Byzantine behavior at a reasonable cost compared to vanilla TensorFlow deployment.\\

\textit{Weaknesses:} GuanYu's assumption of a minimal difference between models on correct parameter servers cannot be applied in some cases.

\paragraph{\textbf{Trimmed mean}}
TianXiang et al. \cite{ tianxiang2019aggregation} proposed a new aggregation rule, Trimmed mean, based on the mean aggregation rule. Trimmed mean indicates the optimization direction, and in the dataset, it refers to the data pruning, and the data with the most significant difference are eliminated. Afterward, the authors present the Trimmed mean's dimensional Byzantine resilience theorem, which shows that the Trimmed mean can satisfy the dimensional Byzantine resilience.

Trimmed mean achieves a nearly linear time complexity O(dnlogn). When the authors compare the Trimmed mean with the set of aggregation rules, they prove its robustness, where it can still converge to the optimal global solution.\\

\textit{Weaknesses:} Trimmed mean can be applicable only in the convex environment.

\paragraph{\textbf{FABA}}
Xia et al. \cite{xia2019faba} proposed FABA (a fast aggregation algorithm against Byzantine attacks), which helps the participating workers obtain correct gradients and avoid outliers in the distributed training. FABA is based on four essential things in the input phase:
\begin{itemize}
\item The set of workers where it computed the gradients.
\item The weights.
\item The learning rate.
\item The Byzantine workers assumed proportion.
\end{itemize}

Considering that $G_{mean}$ is the mean of all the gradients uploaded from workers, the FABA algorithm is executed as follows:

\begin{enumerate}
\item If the number of attack gradients $(k)$ is strictly less than the multiplication of all the gradients uploaded from the workers and the supposed proportion of Byzantine workers, FABA continues its execution; otherwise, it goes to step $5$.
\item Calculate $G_{mean}$ as $g0$.
\item For each gradient in $G_{mean}$, calculate the difference between $g0$ and it. Remove the one with the biggest difference from $G_{mean}$.
\item In this step, FABA adds $1$ to the number of attack gradients, $ k = k + 1$, and returns to step $1$.
\item Calculate the $G_{mean}$ as an aggregation result at a specific time.
\item Update the weights and return them to each worker. 
\end{enumerate}

FABA is easy to implement; in the presence of Byzantine workers, FABA is fast to converge and can tune performance adaptively. It can achieve the same performance and accuracy as the non-Byzantine case. Compared to the previous algorithms of FABA, it shows high efficiency. The authors also show that the aggregation gradients of the FABA algorithm are close to the gradients calculated by honest workers. They proved that these true gradients bounded the gradients in the aggregation moments.\\

\textit{Weaknesses:} FABA's satisfaction condition requires that all gradients from honest workers are collected together while their differences are small. To satisfy this requirement, workers must receive consistently selected datasets, and batch sizes should not be too small. However, this is not the case with each worker having their own private dataset and unsure how the datasets are distributed.

\paragraph{\textbf{BRIDGE}}
Yang and Bajwa \cite{yang2019bridge} presented BRIDGE (Byzantine-resilient decentralized gradient descent), a decentralized learning method, which is a particular case of the distributed gradient descent (DGD) algorithm \cite{nedic2009distributed}. Compared to the DGD algorithm, the BRIDGE algorithm has a screening step before each update, making it Byzantine resilient. The screening method called coordinate-wise trimmed mean eliminates all \textit{b} values (largest and smallest) in each dimension, where \textit{b} presents the bound of Byzantine nodes the algorithm can tolerate. Therefore, the authors combine dimension-wise trimmed mean with decentralized gradient descent to solve the problem of decentralized vector-valued learning under Byzantine settings. 

The authors show the efficiency of the proposed algorithm in dealing with decentralized vector-valued learning issues, provide theoretical assurance for the strongly convex problems, and show by numerical experiments the utility of BRIDGE on non-convex learning problems.\\

\textit{Weaknesses:} BRIDGE is vulnerable to some Byzantine attacks as a result of its distance-based design strategy.

\paragraph{\textbf{LICM-SGD}}
Yang et al. \cite{yang2019byzantine} proposed LICM-SGD (Lipschitz-inspired coordinate-wise median) to mitigate Byzantine attacks.
As this intuition: "benign workers should generate stochastic gradients closely following the Lipschitz characteristics of the true gradients." the proposed LICM-SGD algorithm is inspired. Each true worker makes the stochastic gradient calculation based on the mini-batch and sends the result to the PS in the training process. The PS receives an arbitrary value in the case of a Byzantine worker. When the PS gets all stochastic gradients from the set of workers, it computes the coordinate-wise median. In the last step, and before that, LICM-SGD updates the parameter vector, which makes the selection of true gradients based on Lipschitz characteristics. 

LICM-SGD features several advantages:
\begin{itemize} 
\item In the non-convex setting, LICM-SGD converges certainly to the stationary region, where it can face up to half of the Byzantine workers.
\item For the practical implementations, LICM-SGD is interesting, where it needs no information on the attackers' number and the Lipschitz constant.
\item LICM-SGD has an optimal computational time complexity, as the standard SGD's time complexity under no attacks.
\item As its low computational time complexity, LICM-SGD achieves a faster running time.\\
 \end{itemize} 

\textit{Weaknesses:}
The LICM-SGD method is based only on the i.i.d data.

\paragraph{\textbf{Gradient Filter CGC}}
Gupta and Vaidya \cite{gupta2019byzantine_b} proposed the gradient-filter CGC $(Comparative$ $Gradient$ $Clipping)$ to solve the linear regression problem. Gradient-filter CGC eliminates Byzantine faulty workers' harmful effects and enhances gradient aggregation in the parallelized SGD process. With the assumption of $ n > 2f $, the CGC works as follows, in each iteration, the algorithm work on clipping the set of the $f$ largest gradients to the $2$-norm, which is equal to the $2$-norm of, either the $(f +1)$-th largest gradient, or the $(n - f )$-th smallest gradient, and the rest of gradients remain unchanged. The resulting gradients are averaged in the next update of the current estimate.

An advantage of this approach is that a good estimate of the regression parameter \textit{w} was obtained by the parallelized SGD, where the number of defective Byzantine workers is less than half the number of participating workers. Also, in the case of $n = f$, the upper limit of the guarantee estimation error increases only linearly.\\

\textit{Weaknesses:}
This approach's synchronous master-workers system is still subject to a single point of failure.

\paragraph{\textbf{LIUBEI}}
El Mhamdi et al. \cite{mhamdi2019fast} presented LIUBEI, built using TensorFlow. LIUBEI, compared to standard non-Byzantine resilient algorithms, is a resilient Byzantine ML algorithm that does not trust any individual network component. This mechanism is based on the gradient aggregation rules (GARs), network synchrony, and Lipschitz continuity of the loss function. It replicates the parameter server on multiple machines to tolerate the Byzantine workers. LIUBEI introduces a filtering mechanism that workers use to avoid extracting a suspected model from a parameter server. This mechanism comprises two components: the Lipschitz filter and the models filter. The authors show that utilizing the Lipschitz filter without the model filter and vice versa does not ensure Byzantine resilience. The Lipschitz filter component limits the models' growing gradients, and the models filter limits the distance between models on the correct servers in two successive iterations. This bounded distance is guaranteed using the $scatter / gather$ proposed protocol, which works iteratively in two main phases ($scatter$ and $gather$).

In the scattering phase:
\begin{enumerate}
\item The set of parameter servers is working on its local date.
\item There is no communication among the parameter servers.
\item Each worker in each iteration communicates with $f_{ps} + 1$ server to pull the model ($f_{ps}$: the set of servers which can be Byzantine).
\item Such a step continues for T learning iterations.
\end{enumerate}
In the gather phase:
\begin{enumerate}
\item To aggregate the models, the set of parameter servers communicate together.
\item By the set of workers, LIUBEI aggregates models from all servers.
\item Such a step is carried out for a single learning iteration.
\end{enumerate}

Theoretically, the authors prove that LIUBEI is a Byzantine resilience on both sides, servers and workers. LIUBEI guarantees convergence, an accuracy loss of around 5\%, and about 24\% convergence compared to vanilla TensorFlow. Also, the throughput gain is 70\% compared to the Byzantine–resilient ML algorithm, which supposes an asynchrony of the network.\\

\textit{Weaknesses:} LIUBEI still requires more study in the context of enhancing convergence by using the gather step more often, as well as the overhead of communication in this step, which depends on data and model.

\paragraph{\textbf{Stochastic-Sign SGD}}
Jin et al. \cite{jin2020stochastic} are based on SIGNSGD \cite{bernstein2018signsgd} and used stochastic-sign-based gradient compressors to propose Stochastic-Sign SGD. Additionally, they used another Error-Feedback of Stochastic-Sign SGD to enhance the federated learning \cite{mcmahan2017communication} performance. At first, the authors offer two stochastic compressors, sto-sign, and dp-sign. The sto-sign compressor extends to Sto-SIGNSGD. The workers in this scheme implement two-level stochastic quantization and send out the signs of quantized results as an alternative to directly passing signs of gradients. Also, to ensure robustness and 1-bit compressed communication between server and worker, Sto-SIGNSGD used the majority vote rule in gradient aggregation. Then, a dp-sign differentially private stochastic compressor is proposed; it is extended to DP-SIGNSGD to improve privacy and accuracy and preserve communication efficiency. Moreover, the authors developed the proposed algorithm for the Error-Feedback Stochastic-Sign SGD scheme, which exhibited the Stochastic-Sign SGD error feedback. In majority vote operations, errors can occur, which can be tracked and compensated by the server using the Error-Feedback Stochastic-Sign SGD scheme.

The proposed approach can deal with heterogeneous data distribution. The authors prove its convergence with the same rate as SIGNSGD in the situation of homogeneous data distribution. In addition, they theoretically guaranteed the Byzantine resilience of the Stochastic-Sign SGD approach.\\

\textit{Weaknesses:} The model's accuracy is reduced as a result of differential privacy mechanism.

\begin{table}[h]
   \caption{Summary of some recent results concerning synchronous BFT in DML.($\checkmark$): means that the type of attack it takes into consideration in the analyzed approach, ($\times$) means the reverse of $\checkmark$, and ($-$) means that, to the best of our knowledge, information is not supplied in the original resource or other studies. $n$: The total number of worker machines. $f$: Byzantine workers. $d$: The model dimension.}
  \label{tab:attackSum}
  
  \begin{minipage}{\columnwidth}
  \small
  \begin{center}
  \begin{tabularx}{\linewidth}{llllllll}
 %\begin{tabular}{p{2 cm} l p{1.9 cm} l p{1.65 cm} lllll}

%\toprule
 %   Approach&Time complexity&Condition on Byzantine workers number&\parbox[t]{1.3 cm}{Gaussian attack}&\parbox[t]{0.9 cm}{Omniscient attack}&\parbox[t]{1.3 cm}{Bit-ﬂip attack}&\parbox[t]{1.3 cm}{Gambler attack}&\parbox[t]{1.3 cm}{sign-ﬂipping attack}\\
   % \midrule
    \toprule
    Approach&\parbox[t]{2cm}{Time complexity}&\parbox[t]{2.5cm}{Condition on Byzantine workers number}&\parbox[t]{1cm}{Gaussian attack}&\parbox[t]{1.5cm}{Omniscient attack}&\parbox[t]{1cm}{Bit-ﬂip attack}&\parbox[t]{1cm}{Gambler attack}&\parbox[t]{1cm}{Sign-ﬂipping attack}\\
    \midrule
    
      Krum, Multi-krum&$ O(n^2 d) $&$n > 2 f + 2$ &$\checkmark$&$\checkmark$&$\times$&$\times$&$\times$\\
   GBT-SGD(GeoMed)&$ O (nd log^3 1/\varepsilon)$&$n>2f$&$\checkmark$&$\checkmark$&$\checkmark$&$\checkmark$&$\times$\\
   GBT-SGD(MarMed)&$O(dn log n)$&$n>2f$&$\checkmark$&$\checkmark$&$\checkmark$&$\checkmark$&$\times$\\
   GBT-SGD(MeaMed)&$O(dn)$&$n>2f$&$\checkmark$&$\checkmark$&$\checkmark$&$\checkmark$&$\times$\\
   Phocas and Trmean&$O(dn)$&$n>2f$&$\checkmark$&$\checkmark$&$\checkmark$&$\checkmark$&$\times$\\
   Zeno&$O(nd)$&unbounded number&$\times$&$\checkmark$&$\times$&$\times$&$\checkmark$\\
   RSA&$-$&unbounded number&$\times$&$\times$&$\times$&$\times$&$\checkmark$\\
   SIGNSGD&$-$&$n>2f $&$\checkmark$&$\times$&$\times$&$\times$&$\times$\\
   Trimmed mean&$O(dnlogn)$&$n > 2f $&$\checkmark$&$\checkmark$&$\times$&$\times$&$\times$\\
   FABA&$-$&$f .n, f<0.5$&$\checkmark$&$\times$&$\times$&$\times$&$\times$\\
\bottomrule
%\end{tabular}
\end{tabularx}
\end{center}
\bigskip
\end{minipage}
\end{table}

\item \textit{Byzantine fault tolerance approaches based on coding schemes:}

\paragraph{\textbf{DRACO}}
Chen et al. \cite{chen2018draco} Present the first work, which deals with Byzantine workers using a specific encoding scheme with redundant gradient calculations. More clearly, DRACO treats the problem of adversarial compute nodes based on coding theory to achieve robust distributed training. DRACO eliminates the adversarial update's effects by the parameter server; this one uses the redundant gradients evaluated by each compute node. Firstly, the parameter server detects the arbitrarily malicious compute nodes; after that, by the correct nodes shipped the gradient updates, the parameter server recovers the correct gradient average; Hence, DRACO removes the Byzantine values. 

DRACO applies to many training algorithms (the authors focus on mini-batch SGD) and is robust to adversarial compute nodes. Maintaining the correct update rule, DRACO achieves the theoretical lower bound of redundancy, which is necessary to resist adversaries.\\

\textit{Weaknesses:} This framework's main limitations include its defense against a few Byzantine workers and the reduced performance of inexact models when updated. 

\paragraph{\textbf{DETOX}}
Rajput et al. \cite{rajput2019detox} present a Byzantine-resilient distributed training framework, DETOX. This framework applies to the parameter server architecture and is based on two steps. Firstly it uses the algorithm's redundancy to filter the majority of the Byzantine gradients. In the second step, DETOX performs a hierarchical robust aggregation method, where it divides a few groups by partitioning filtered gradients and aggregating them. After that, and to minimize the remaining traces' effect of the original Byzantine gradients, it uses any robust aggregator to the averaged gradients. 

DETOX has good scalability, flexibility, and nearly linear complexity. DETOX improves its efficiency and robustness when combined with any previous robust aggregation rule. Also, when compared to the vanilla implementations of Byzantine-robust aggregation, DETOX increases accuracy to 40\% in the event of heavy Byzantine attacks.\\

\textit{Weaknesses:} DETOX needs the parameter server as an owner to partition the data among the workers, which breaks the data privacy.

\paragraph{\textbf{RRR-BFT}}
Gupta and Vaidya \cite{gupta2019randomized} proposed two schemes (deterministic and randomized) for guaranteeing exact fault-tolerance \footnote{ The authors in \cite{gupta2019randomized}expressed that despite the presence of Byzantine workers, a parallelized-SGD method has exact fault-tolerance if the master asymptotically converges to a minimum point w (w is a minimum point of the average loss evaluated for the data points.) exactly.} in the parallelized-SGD method. These schemes are based on the "computation efficiency" \footnote{ The computation efficiency of the coding scheme was presented by the ratio of two types of numbers. The first is the gradient's number used for parameter update, and the second is the gradient's number computed in total by the workers.} of a coding scheme, and the authors considered the case where $f (< n=2) $ of the workers are Byzantine faulty. If Byzantine workers could send failure data to the master, the proposed approach used the reactive redundancy mechanism to identify and isolate this type of worker.

In the deterministic scheme:
\begin{enumerate}
\item The master uses a symbol (fault detection code) at each iteration.
\item The master used the reactive redundancy mechanism when it detected any fault(s) to correct the faults and identify the Byzantine worker that caused the fault(s).
\end{enumerate} 

In the randomized scheme:
\begin{enumerate}
\item This scheme is used to improve upon the deterministic scheme's computation efficiency. Instead of using all iterations, the fault-detection codes used by the master are chosen randomly in the intermittent iterations.
\end{enumerate}

The main advantage of this work is the exact fault tolerance and favorable computation efficiency. The master can optimize the compromise between the expected calculation efficiency and the parallelized learning algorithm's convergence rate.\\

\textit{Weaknesses:}
RRR-BFT is still susceptible to a single point of failure.\\

\item \textit{Byzantine fault tolerance approaches based on Blockchain:}

\paragraph{\textbf{DeepChain}}
Weng et al. \cite{weng2019deepchain} proposed DeepChain for privacy-preserving deep learning training to provide auditability and fairness; this approach is a Blockchain-based technology and a decentralized mechanism to solve the perturbation problem in the collaborative training process, which the malicious participants cause. 

Before explaining the DeepChain approach, we provide the DeepChain terms used by the authors as follows:
\begin{itemize}
  
\item The worker is analogous to Bitcoin miners.
\item Round describes the process of building a new block.
\item DeepCoin is a particular type of asset that DeepChain will generate as incentives.
\item Party is the same entity represented in the traditional distributed deep learning model, with equivalent requirements; however unable to execute the entire training task without assistance due to resource restrictions such as limited compute capacity or restricted data.
\item Trading is when a party receives her local gradients, and she transmits them to DeepChain using a smart contract called a trading contract. These contracts can be downloaded to process by the worker.
  
\end{itemize}
DeepChain is an incentive mechanism that consists of five components: 
\begin{itemize}
    \item \textbf{DeepChain bootstrapping :} it comprises two steps, DeepCoin distribution and genesis block generation. Initially, DeepCoins are allocated equally among parties and workers. Following that, each DeepCoin has initial transactions that record ownership statements on a genesis block generated at round 0.
    \item \textbf{incentive mechanism :} it drives participants to participate actively and honestly in collaborative training tasks, intending to reward or penalize participants based on their contribution.

\item \textbf{asset statement :} as part of her deep learning task, the party needs to state her assets in order to find cooperators and achieve the task.

\item \textbf{collaborative training :} parties with similar deep learning tasks can form a collaborative group based on their stated assets. It is then decided which information is needed to train the deep learning model securely. Gradients are then collected via \textit{Trading Contracts}, in which parties trade their gradients iteratively through a manager chosen from cooperative parties. All traders encrypt their trading gradients honestly and attach the correct encryption proofs that confirm confidentiality and auditability. Finally, workers process transactions by adding up gradients, which are reported to the \textit{Processing Contract}, which verifies that the computation results are correct, so that group parameters can be updated.
With DeepChain, two security mechanisms aid in enhancing fairness in collaborative training. A trusted time clock mechanism forces operations in a contract to be completed before a particular period. Secondly, there is the secure monetary penalty mechanism, consisting of gradient collection and collaborative decryption functions. DeepChain applies the monetary penalty mechanism, revoking the pre-frozen deposit of dishonest participants and reallocating it to the honest participants. Through this mechanism, dishonest participants' pre-frozen deposits are revoked and reallocated to honest participants. Fairness can be achieved by not penalizing honest participants that behaved punctually and correctly, and they will be compensated if dishonest participants take part.

\item \textbf{consensus protocol :} is used to ensure that all participants can make a consensus upon an event.
\end{itemize}

DeepChain proves its performance in terms of confidentiality, auditability, and fairness.\\

\textit{Weaknesses:} the security challenge should be re-defined if DeepChain expands potential models' value to transfer learning.

\paragraph{\textbf{LearningChain}}
Chen et al. \cite{chen2018machine} discussed privacy and security problems in the linear/ non-linear learning models. The authors designed the decentralized SGD algorithm to learn a general predictive model over the Blockchain and proposed a framework based on Blockchain technology called LearningChain. LearningChain is a decentralized, privacy-preserving, and secure machine learning system. Explicitly, it consists of three processes. The first one is the \textit{Blockchain Initialization}; in this process, the computing nodes and data holders establish a connection and construct a fully decentralized peer-to-peer network. The second process is the \textit{Local Gradient Computation}: based on an actual global model and common loss function, the data holders create pseudo-identities and calculate the local gradients. In the following, to perturb their local gradients, they use a differential privacy scheme. Finally, the messages are broadcast to all compute nodes after the computed local gradients are encapsulated with other related information. The third process is the \textit{Global Gradient Aggregation}: finding a solution to a mathematical puzzle is a condition of the computing nodes that compete for permission to add a new block to the chain. The aggregation scheme of tolerance to Byzantine attacks applies when a compute node wins the game to aggregate the local gradients. Furthermore, it calculates the global gradient to update the model parameters \textit{w}. Finally, the associated information adds a newly created block to the chain.
LearningChain can train a global predictive model by repetitively performing local gradient computation and global gradient aggregation by participants' collaboration in the network while keeping their own data privacy.

The authors develop a differential mechanism based on privacy to protect each party's data privacy. Moreover, to protect the system and defend against any possible Byzantine attacks, they come up with a $l$-nearest aggregation algorithm. In the implementation phase, the authors prove the LearningChain efficiency and effectiveness on Ethereum, also by its extension, LearningChainEx applied to all types of Blockchain.\\

\textit{Weaknesses:} LearningChain used the proof-of-work consensus, which requires substantial energy for every transaction. However, the energy efficiency has not been verified.
\end{enumerate}

\subsubsection{Synchronous training Byzantine fault tolerance approaches in other machine learning optimization methods}
\begin{enumerate}
\item \textit{BFT approaches based filtering schemes:}
\paragraph{\textbf{DSML-Adversarial}}
Chen et al. \cite{chen2017distributed} considered the decentralized systems to solve the problem of distributed statistical learning with existing Byzantine attacks. The authors proposed a variant of standard gradient descent based on the geometric median aggregation rule. They addressed the challenges facing federated learning (security, small local datasets versus high model complexity, communication constraints) and proposed their own Byzantine gradient descent method. The latter is executed as follows: 

The parameter server partitions the set of working machines $n$ into $k$ batches. During each iteration, gradients are grouped in non-superimposed batches to increase the resemblance between Byzantine-free batches and prevent the Byzantine machines' interruption. After that, it computes the mean of each local received gradient in each batch. Then, it computes the geometric median of the "k" batch means. Finally, it performs the gradient descent update with the aggregated gradient.

In addition that DSML-Adversarial can tolerate Byzantine failures; with a small number of local data, it can learn a very complex model and achieve fast exponential convergence.\\

\textit{Weaknesses:} the training data does not distribute equally, the irregular availability of mobile phones, and the algorithm does not adapt to the asynchronous parameter. In addition, federated learning allows users to store their training data locally, giving them control over their data. However, it is necessary to extract this data to train a high-quality model. As a result, the proposed algorithm must specify the minimum confidentiality level sacrificed in the federated learning paradigm.

\paragraph{\textbf{ByzRDL}}
Yin et al. \cite{yin2018byzantine} made a study around the BRDSL (Byzantine-robust distributed statistical learning) algorithms to solve the optimal statistical performance problem. The authors, based on the standard statistical setting of empirical risk minimization (ERM). They proposed three algorithms to simultaneously achieve the $Statistical$ $optimality$ and $Communication$ $efficiency$. Two distributed gradient descent algorithms, where the first one is based on the coordinate-wise median, and the second is based on the coordinate-wise trimmed mean. The latest is the median-based robust algorithm, which only needs one communication round. The authors proved the statistical error rates for the three types of population loss functions (strongly convex, non-strongly convex, and non-convex) and the algorithms’ robustness against Byzantine failures. They verified the optimal error rates of the proposed algorithms: Under mild conditions, the median-based GD and Trimmed-mean-based GD achieve the order optimal for strongly convex loss. The median-based one-round algorithm achieves the optimal rate for strongly convex quadratic losses, similarly to the optimal error rate achieved by the robust distributed gradient descent algorithms. When the Byzantine workers are up to half of all workers, ByzRDL achieves convergence.\\

\textit{Weaknesses:} ByzRDL does not converge when the Byzantine workers represent more than half of all workers.

\paragraph{\textbf{DGDAlgorithm}}
Cao et al. \cite{cao2019distributed} proposed DGDAlgorithm Robust (Distributed Gradient Descent Algorithm Robust), an algorithm to solve the problem of arbitrary Byzantine attackers falsifying data. Based on the gradient compute, the DGDAlgorithm tests the data sent by workers. In the first step, the algorithm asks the parameter server to select a small set of clean data randomly. Then, a noisy gradient is computed at each iteration $t$ by selecting a small dataset. After comparing the receiving gradients from the workers with the noisy local gradients, the parameter server accepts the receiving gradients if the difference is within a threshold. Despite the presence of Byzantine workers, the authors proved that the proposed algorithm converged to the optimal value.\\

\textit{Weaknesses:} DGDAlgorithm remains inapplicable on the non-convex cost function, vulnerable to the single point of failure due to the collection of gradients required on the parameter server. The threshold requires a manual adjustment.

\paragraph{\textbf{Securing-DML}}
Su and Xu \cite{su2018securing} proposed a secured variant of the classical gradient descent method to tolerate the Byzantine workers in distributed/decentralized learning systems. The authors considered the full gradient descent method and the $ d $–dimensional model. The proposed method asked each non-defective worker to calculate the gradient based on the entire local sample. Then it used the proposed filtering approach \cite{steinhardt2018resilience} for the problems of robust mean estimation to aggregate the gradients reported by the external workers robustly in the distributed statistical learning setting. Iterates and the associated gradients are strongly interdependent. However, this interdependence can not be speciﬁed since Byzantine workers might act arbitrarily \cite{su2018securing}. According to the authors, to deal with this unspecified dependency, the sample covariance matrix's gradient concentration should be set uniformly at all possible parameters of the model.
Charikar et al. \cite{charikar2017learning} derived the same uniform convergence concentration of the sample covariance matrix under the assumption of sub-Gaussian gradients. They used sub-exponential gradients instead of sub-Gaussian gradients in the simplest linear regression example. Therefore, Su and Xu \cite{su2018securing} developed a novel concentration inequality for sample covariance matrices of sub-exponential distributions.

Up to a constant fraction of Byzantine workers can be tolerated by the proposed method with $ O(logN) $ ($ N $: total number of data points distributed across $n$ workers) communication rounds, and it converges to a statistical estimation error on the order of $ O(\sqrt{f/N} + \sqrt{d/N}) $.\\

\textit{Weaknesses:} Securing-DML remains inapplicable to the non-convex cost function.

\paragraph{\textbf{ByzantinePGD}}
Yin et al. \cite{yin2019defending} developed a ByzantinePGD (Byzantine Perturbed Gradient Descent) algorithm to solve the security problems in large-scale distributed learning. These security problems may be due to the presence of Byzantine workers with a non-convexity loss function.

ByzantinePGD is based on the PGD algorithm described in \cite{jin2017escape} and employs the following strategy to combat Byzantine workers: 

\begin{enumerate}
    \item It uses the three robust aggregation rules (median \cite{kogler2016efficient}, trimmed mean, iterative filtering \cite{diakonikolas2019robust, diakonikolas2017being, steinhardt2018resilience}), where the authors present GradAGG and ValueAGG subroutines that collect gradients and function values robustly from workers. These subroutines use the terminology of inexact oracle to formalize the guaranteed accuracy.
    \item It escapes saddle points by random perturbation in multiple rounds.
\end{enumerate}
The authors proved that ByzantinePGD converges to an approximate local minimizer and escapes the saddle points with the existence of the Byzantine machine and the nonconvex function.\\

\textit{Weaknesses:}  ByzantinePGD is still near-optimal in high-dimensional settings.

\paragraph{\textbf{ByRDiE}}
Yang and Bajwa \cite{yang2019byrdie} proposed the Byzantine resilient distributed coordinate descent (ByRDiE) algorithm to deal with Byzantine failures in decentralized settings. The authors proposed two main contributions, where the algorithmic aspects of ByRDiE are leveraged on two separate lines of prior work. The first one, with Byzantine failures, an inexact solution is possible in certain types of scalar-valued distributed optimization issues \cite{su2015fault, su2016fault}. Second, break down the optimization problems of vector-valued into a sequence of scalar-valued problems with coordinate descent algorithms \cite{wright2015coordinate}. This phase represents the proposed work's first main contribution, and the theoretical analysis represents the second main contribution. In particular, ByRDiE divides the empirical risk minimization (ERM) problem into P one-dimensional sub-problems using coordinate descent. Then, it uses the Byzantine resilient approach presented in \cite{su2015fault} to solve each scalar-valued sub-problem.

The theoretical results ensure the ByRDiE's ability to minimize the statistical risk in a fully distributed environment (decentralized settings). The authors also prove the resilience of ByRDiE to cope with Byzantine network failures for distributed convex and non-convex learning tasks.\\

\textit{Weaknesses:} ByRDiE still needs to strengthen its results regarding near-safe convergence and derivation of explicit convergence rates. Moreover, ByRDiE is still vulnerable to Byzantine attacks and requires additional theoretical studies on functions of constant step size and non-convex risk functions.

\paragraph{\textbf{SVRG-AdversarialML}}
Bulusu et al. \cite{bulusu2019convex} adopted the stochastic variance reduced gradient (SVRG) optimization method to solve the problem of Byzantine workers in distributed settings. Considering $n$ workers and the finite-sum issue in the convex situation, the authors proposed a resilient variation of SVRG. If the objective function $F$ is convex with the existing $f$-fraction of Byzantine workers, the proposed method serves to:
\begin{enumerate}
\item Firstly, finds an optimal point $x$ within $T$ iterations $ ( T= \tilde{O}(\frac{1}{\gamma}+\frac{1}{n\gamma^2}+\frac{f^2}{\gamma^2}))$, with $\gamma$ denotes the discount factor. 
\item Secondly, $\mathbb{E}[F(x)- F(x^*)] $ must be satisfied.
\end{enumerate}
To achieve the SVRG's resilience, the authors calculate the average of every worker's intermediate gradients over each iteration, as well as the median across all workers.

The authors proved the convergence of their proposed variant of the SVRG algorithm in the convex case and its robustness in adversarial settings (adversarial attacks and reduced variance).\\ 

\textit{Weaknesses:} SVRG-AdversarialML algorithm remains inapplicable in non-convex and strongly convex cases.

\paragraph{\textbf{DRSL-Byzantine Mirror Descent}}
Ding et al. \cite{ding2019distributed} developed a distributed robust learning algorithm based on mirror descent \cite{bubeck2015convex}. This algorithm solves the distributed statistical learning problem, safeguards against adversarial workers, and shares fault information in a high-dimensional learning system. DRSL-Byzantine Mirror Descent uses the median aggregation rule to deal against $ f $ Byzantine workers from $ n $ workers. The authors defined two different formulations of the sum of the population $  B_T^i $ and $ A_T^i$, and described the robust variant of mirror descent to the Byzantine setting as follows:
\begin{enumerate}
	\item For each honest worker $ (i) $, by the martingale concentration, $ B_T^i $ will concentrate in each iteration $t \in [T] $, on the population sum $B_T^\star$ of the gradient $ \bigtriangledown_t$, with an utmost deviation of $ \sqrt{T}$.
	\item The authors defined a set of workers as Byzantine if they are too far from the mean.
	\item With a small deviation in terms of $ B_t^i$, some Byzantine workers can be hidden.
	\item Hence, the authors identified such Byzantine workers by the martingale concentration for $ A_T^i$. In the case where $ A_T^i $ is too far away from the population sum $A_T^\star$, the authors consider them Byzantine workers.
	\item The last step is to remove the set of Byzantine workers and put the rest in the set $\Omega_{t}$ (estimated collection of honest workers at iteration $t$).
\end{enumerate}

The proposed method is robust against the Byzantine workers for $ f  \in [0.1⁄2)$, also it achieves a statistical error bound $ O(1⁄\sqrt{nT}+ f⁄\sqrt{T}) $ for $T$ iterations in the convex and smooth objective function. This result is similar to the known statistical error for the Byzantine stochastic gradient in the Euclidean space framework. The Byzantine mirror descent algorithm depends on the dimension to the bound scales with the dual norm of the gradient. In particular, its features depend logarithmically on the problem dimension $d$ for probability simplex.\\

\textit{Weaknesses:} DRSL-Byzantine Mirror Descent is vulnerable to a single point of failure.

\paragraph{\textbf{MOZI}}
Guo et al. \cite{guo2020towards} proposed MOZI, a Byzantine fault tolerance algorithm in decentralized learning systems. Using a uniform Byzantine-resilient aggregation rule, MOZI allows the honest node to train a correct model with an arbitrary number of faulty nodes under powerful Byzantine attacks. In each training iteration, the uniform Byzantine-resilient aggregation rule chooses useful parameter updates and filters out harmful ones. MOZI detects the Byzantine parameters thanks to distance-based and performance-based strategies. These strategies are used to achieve two goals: 
\begin{enumerate}
	\item The distance-based strategy seeks to decrease the scope of candidate nodes in order to achieve higher performance. 
    \item The performance-based strategy is utilized for lifting all faulty nodes to provide robust Byzantine fault tolerance.
\end{enumerate}
The authors adopted a distance-based selection phase to counter sophisticated Byzantine attacks that can modify the baseline value. Instead of the mean or median value of received incorrect parameters, each honest node uses its own parameters as the baseline. By comparing each honest node's estimate to the Euclidean distance of each neighbor node's estimate, a distance-based strategy is used to select a candidate pool from among the prospective honest nodes. As a result, the quality of chosen nodes was enhanced.
In the performance-based selection phase, each honest node tests the chosen parameters' performance using its training samples. Hence, it can alleviate the unreasonable assumption in performance-based strategies of the server's availability of validation datasets in centralized systems. More specifically, each honest node reuses the training sample as validation data to test the performance of each estimate. To obtain the final update value, it averages the loss values of the selected estimates that are smaller than its own estimate.

The authors proved the strong convergence of MOZI in their theoretical analysis. Experimental results demonstrate that MOZI can handle both sophisticated and simple Byzantine attacks with little computational cost.\\

\textit{Weaknesses:} Mozi is intended only for supervised learning in i.i.d.

\paragraph{\textbf{Distributed Momentum}}
El-Mhamdi et al. \cite{el2020distributed} proved the robustness and the advantages of using Momentum on the workers' side (considering that Momentum is generally used on the side of the server and Byzantine-resilient aggregation rules are not linear) to guarantee "quality gradient". 

Firstly, Momentum decreases the variance-standard ratio of computing the gradient at the server level and expands the Byzantine resilient aggregation rules; second, its resilience influences distributed SGD. Furthermore, the authors demonstrated that a combination of Momentum (on the worker side) and standard defense mechanisms (KRUM\cite{blanchard2017machine}, MEDIAN\cite{xie2018generalized}, BULYAN \cite{el2018hidden}) could be sufficient to defend against the two presented attacks in \cite{baruch2019little, xie2020fall}.\\

\textit{Weaknesses:} this work still requires further study and analysis of the interplay between the dynamics of the two types of optimization approaches (stale gradients and Momentum).\\

\item \textit{BFT approaches based coding schemes:}

\paragraph{\textbf{BRDO based Data Encoding}}
Data et al. \cite{data2020data} studied distributed optimization in the master-worker architecture with the presence of Byzantine adversaries. The authors focused on two iterative algorithms: Proximal Gradient Descent (PGD) and Coordinate Descent (CD). Gradient descent is a special case of both PGD and CD. The first algorithm is used in the data-parallel setting, while the second is used in the model-parallelism setting.
The proposed algorithm is based on data encoding/decoding where: 
\begin{enumerate}
\item In the data encoding, the algorithm uses the sparse encoding matrices in the data used by the set of worker nodes.
\item An efficient decoding scheme using the error correction over real numbers \cite{candes2005decoding} has been developed at the master node to process the workers' inputs.
 \end{enumerate}
In the case of gradient descent, the developed scheme is used to calculate the correct gradient, and in the case of coordinate descent, it facilitates the calculation at working nodes. The encoding scheme extends efficiently to the data streaming model and achieves stochastic gradient descent (SGD) Byzantine-resilient.

The algorithm provided by the authors can tolerate an optimal number of corrupt worker nodes. Furthermore, it makes no assumptions about the statistical probability distributions of the data or the Byzantine attack models. It can also tolerate up to a third of corrupt worker nodes, with constant overhead for computational, communication, and storage requirements, in contrast to the distributed PGD/CD process, which does not provide any security.\\

\textit{Weaknesses:} Because this method presupposes i.i.d data, it falls short in the scenario of federated learning, when data encoding across several nodes is unfeasible.

\item \textit{BFT approaches based Blockchain:}
\paragraph{\textbf{TCLearn}}
Lugan et al. \cite{lugan2019secure} proposed scalable security architectures and Trusted Coalitions for Distributed Learning (TCLearn). TCLearn used suitable encryption and Blockchain mechanisms to protect data privacy, ensure a consistent sequence of iterative learning, and fairly share the learned model across coalition members.
In the proposed approach:
\begin{enumerate}
\item The coalition members shared a CNN model, which was optimized in an iterative sequence.
\item Each coalition member updates the shared model with new batches of local data in a sequential manner.
\item Validated the shared model via a process involving coalition members, then stored on the Blockchain.
\item Blockchain has provided an immutable ledger that allows retrieving each step of the model evolution.
\item The authors proposed a consensus mechanism Federated Byzantine Agreement (FBA). The proposed TCLearn relied on FBA to provide an iterative certification procedure for each stage of learning the shared model, control the quality of the updated CNN model, and avoid any inadequate training to add corrupted increments to the model. As a result, the proposed FBA verified and validated the addition of a new block to the chain, as well as its integrity.
\end{enumerate}
According to the common rules of the coalition, the authors have developed three methods that correspond to three distinct levels of trust:
\begin{itemize}
\item TCLearn-A method: public learned model, in which each coalition member is responsible for ensuring the confidentiality of their own data.
\item TCLearn-B Method: private learned model, where the coalition's members trust each other.
\item TCLearn-C Method: prevent any dishonest behavior from coalition members who do not trust each other.
\end{itemize}
The encryption system and off-chain storage used by the proposed TCLearn architectures enable data privacy preservation and degradation prevention.\\

\textit{Weaknesses:} The proposed approach does not address the costs associated with privacy preservation.

\paragraph{\textbf{SDPP based-Blockchain System}}
Zhao et al. \cite{zhao2019mobile} presented a system based on mobile edge computing, Blockchain technology, and reputation-based crowdsourcing federated learning. The proposed solution used federated learning technology to create an intelligent system to assist internet of things (IoT) device manufacturers in leveraging customer data. This intelligent system forecasts consumers' needs and potential consumption patterns by constructing a machine learning model. Then, the authors used differential privacy to deal with potential adversaries in federated learning training to protect customers' data. Finally, they take advantage of Blockchain technology to enhance the security of model updates.

The proposed system offers several benefits, including high computing power, confidentiality and auditing, and distributed storage. It protects the privacy of consumers' data while ensuring the security of model updates. It incentivizes users to participate in crowdsourcing tasks, allowing manufacturers of IoT smart devices to improve their businesses by forecasting customer consumption behaviors. \\

\textit{Weaknesses:} this system requires real-world IoT device manufacturers testing.

\paragraph{\textbf{BlockDeepNet}}
Rathore et al. \cite{rathore2019blockdeepnet} presented BlockDeepNet, a decentralized extensive data analysis approach that blends Blockchain technology and DL to obtain secure collaborative DL in IoT. As part of the modern IoT network's Big Data analysis process, malicious attackers can exploit large quantities of data. Therefore, the proposed BlockDeepNet system minimizes the possibility of data being negatively manipulated as follows:
\begin{itemize}
\item At the device level, collaborative DL is executed to avoid privacy leaks and obtain adequate data for DL.
\item To ensure confidentiality and integrity of DL collaboration in the IoT, the authors applied Blockchain technology at the edge server level to aggregate local learning models supported at the device level.
\item To validate the effectiveness of the proposed approach, the authors develop a prototype model of BlockDeepNet in real-time scenarios.
\end{itemize}

The BlockDeepNet experimental evaluation proves Big Data analysis tasks' robustness, security, and feasibility in IoT. BlockDeepNet mitigates current issues, such as a single point of failure, IoT device privacy leaks, a lack of valuable data for DL, as well as data poisoning attacks. Therefore, Blockchain operation for DL in IoT achieves more accuracy while maintaining acceptable latency and computational cost.\\ 

\textit{Weaknesses:} BlockDeepNet still faces DL demands for more computing power.
\end{enumerate}
%----------------------------------------------------------------------------
%----------------------------------------------------------------------------
\subsubsection{Asynchronous training Byzantine fault tolerance approaches}

\paragraph{\textbf{Kardam}}
Damaskinos et al. \cite{damaskinos2018asynchronous} Introduced the Kardam SGD algorithm that tolerates the asynchronous Byzantine behavior in ML. this algorithm was based on two essential components (filtering and dampening). Resilience is ensured by the filtering component found on the scalar against $1/3$ Byzantine workers. This filter acts as a self-stabilizer by leveraging the Lipschitzness of cost functions to protect the SGD model from adversarial attacks. Using the frequency filter introduced by the authors, the filtering component prevents Byzantine workers from keeping honest workers from updating the training model. The dampening component is the second, and it bounds the convergence rate using a generic gradient weighting scheme that adjusts to stale information. 

Kardam provides nearly certain convergence despite Byzantine behavior. According to the authors, Kardam is the first SGD algorithm that tolerates asynchronous distributed Byzantine behavior.\\

\textit{Weaknesses:} Kardam demands a stronger restriction, where it can only deal with the number of Byzantine workers, which is less than one-third of all workers. Moreover, this is a weak assumption, as Byzantine workers can make up half of the existent workers in the standard setting. Furthermore, the filtering component may suffer in the context of a frequency filter, which is too harsh for asynchronous systems. Additionally, in terms of Byzantine resilience's cost, Kardam induces a slowdown. 

\paragraph{\textbf{DBT-SGD in the Era of Big Data}}
Jin et al. \cite{jin2019distributed} proposed two asynchronous SGD algorithms to tolerate an arbitrary number of Byzantine workers. The authors removed the need for a shared parameter server. It is assumed that the set of true collaborative workers stores the local model parameter, which may be utilized as the basic truth. With the latter, workers may get and filter the shared model parameters from other co-workers at any time.

The first algorithm corresponds to a known number of Byzantine workers, with an assumed upper-bound $f$ of the number of Byzantine workers. To prevent the Byzantine attack, each worker takes an average of the parameters of the $n - f - 1$ model that are closest to the parameter of the honest worker model; lastly, a gradient descent update is realized based on this average value.

The number of Byzantine workers is unknown in the second algorithm. Firstly, the worker accepts the model parameters that result in the lowest empirical risk based on the worker's evaluation of local training samples. Gradient descent update is then performed by averaging the acceptable model parameters.

The proposed algorithms effectively against all sorts of Byzantine attacks that have been tested and provide convergence.\\

\textit{Weaknesses:} rather than using past information to enhance performance, the proposed algorithms only consider current shared information when deciding whether to accept them or not.
 
\paragraph{\textbf{Zeno++}}
Xie et al. \cite{xie2020zeno++} proposed Zeno ++, a robust, fully asynchronous procedure based on Zeno's Byzantine-tolerant synchronous SGD algorithm \cite{xie2018zeno}. The authors intend to tolerate Byzantine failures on anonymous workers (with the potential for an unlimited number of Byzantine workers).
The training process is at the heart of this approach. Zeno ++ works to estimate the descent of the loss after applying the candidate gradient to the model's parameters; it scores the received gradients and accepts them based on the score. In addition, for efficient computations, a lazy update is proposed. 

Zeno ++ converges well in non-convex problems, and its empirical results indicate it performs better than earlier approaches.\\ 

\textit{Weaknesses:} Zeno++ is not applicable in many contexts, such as federated learning.

\paragraph{\textbf{BASGD}}
Yang and Li \cite{yang2020basgd} proposed for Byzantine Learning (BL) a method called Buffered Asynchronous Stochastic Gradient Descent (BASGD). In asynchronous BL, the server does not have a stored training instance. Therefore, BASGD overcomes this problem by introducing the server buffers B ($0 <B <= n$) used in the gradient aggregation phase and updating the parameters. The BASGD method is based on two components: $buffer$ and $aggregation$  $function$.

In the Buffer component, the training method in BASGD and ASGD is the same, with the exception of a server-side update rule adjustment, as follows:
\begin{enumerate}
\item B buffers $(0 <B <= n)$ are introduced on the server.
\item When the gradient $(g)$ in the server buffer is received from the worker $(_s)$, the parameter is not instantly updated.
\item Each buffer $b$ holds the incoming gradient tentatively, where $b = s$ $mod$ $B$.
\item The new SGD step can only be executed if all buffers have been modified since the last SGD step.
\item It is possible that each buffer $b$ have been received more than one gradient between two iterations.
\item Each buffer stores the average of the gradients received.
\item An update rule will be applied for each buffer $b$.
\item After the update step, all buffers will be reset at the same time.
\end{enumerate}
In the aggregation function component: a more reliable gradient can be aggregated from the $B$ candidate gradients of buffers when the server updates a parameter using the right aggregation function. The latter takes the average of all candidate gradients but satisfies the authors' q-Byzantine Robust (q-BR) condition\footnote{ For an aggregation function, (q-BR) condition illustrates its Byzantine resilience ability quantitatively.}.

BASGD is more general than synchronous BL methods. It can provide the privacy of asynchronous BL because the server does not have a stored training instance. Additionally,  BASGD resists the set of attackers and has the same theoretical convergence rate as vanilla asynchronous SGD (ASGD). Furthermore, it can greatly outperform vanilla ASGD and other ABL baselines in case of error or attack on the workers' part.\\

\textit{Weaknesses:} To validate the BASGD convergence and resilience against attack or error, the authors assumed a limited number of Byzantine workers.

\paragraph{\textbf{ByzSGD}}
El-Mhamdi et al. \cite{el2020genuinely} studied the issue of Byzantine-resilient distributed machine learning in a decentralized architecture. They introduced the ByzSGD algorithm, which aims to tolerate Byzantine failures on both sides of servers (several replicas of the parameter server) and workers in an asynchronous setting. The authors used the synchronous setting to minimize the amount of communicated messages required by ByzSGD in order to increase its performance. ByzSGD is based on three significant schemes:

\begin{enumerate}
\item Scatter/Gather scheme: this scheme limits the maximum drift among models, which is due to no communication among servers during the scattering phase. In the gathering phase, honest servers communicate and use a Distributed Median-based Contraction (DMC) module.  
\item Distributed Median Contraction (DMC): DMC uses the median geometric properties in high-dimensional settings to approximate the honest servers' parameters closer to each other and achieve learning convergence. However, ByzSGD still requires many communication messages, which can be decreased using synchronous settings.
\item Minimum–Diameter Averaging (MDA): to tolerate Byzantine workers, ByzSGD used MDA as a statistically–robust gradient aggregation rule.
\end{enumerate}

ByzSGD provides an optimal solution to the problem of Byzantine parameter servers (1/3) and Byzantine workers (1/3) in the asynchronous setting. It also ensures Byzantine resilience without additional communication rounds, unlike vanilla non-Byzantine alternatives. Additionally, it reduced the number of communication messages by using the synchronous setting.\\

\textit{Weaknesses:} ByzSGD is still not applied to fully decentralized settings and non-iid data.

\paragraph{\textbf{ColLearning Agreement}} 
El-Mhamdi et al. \cite{ el2020collaborative} addressed the Byzantine collaborative learning issue without trusting any node while considering genuine Byzantine resilience. This approach is founded on asynchronous and heterogeneous settings, considering the general case of non–i.i.d. data. Furthermore, it is based on fully decentralized settings, the more general non–convex optimization case, and the standard stochastic gradient descent as an applied scheme. 

According to the authors, the type of agreement known as the averaging agreement is similar to collaborative learning. Each node begins with an initial vector and seeks an approximation approval on a common vector, ensuring that it remains within an average constant of the maximum distance between the main vectors. The authors provide three asynchronous algorithms to the averaging agreement as follows:
\begin{enumerate}
\item The first solution, Minimum–Diameter Averaging (MDA), needs $(n \geq ou \ge 6f + 1)$ to attain the best feasible averaging constant based on the minimum volume ellipsoid asymptotically. Furthermore, MDA's filtering scheme ensures that no Byzantine input among the chosen vectors exchanged among workers could be arbitrarily bad.
\item The second technique, Broadcast-based Averaging Agreement, can provide optimal Byzantine resilience $ (n \geq ou \ge 3f + 1) $ based on the reliable broadcast, despite the requirement for cryptographic signatures and numerous communication cycles.
\item The third solution, Iterated Coordinate–wise Trimmed Mean (ICwTM). Unlike the second, ICwTM does not rely on signatures. In standard form algorithms that do not use signatures, it achieves an optimal Byzantine resilience $ (n \geq ou \ge 4f + 1) $ by using a coordinate-wise trimmed mean.
\end{enumerate}

Despite the asynchronous setting considered in the proposed technique, the authors occasionally discuss the synchronous setting in which they would assume constraints on the communication delays amongst honest workers and their relative quickness. The Byzantine-resilient averaging agreement algorithm achieves a quasi-linear complexity.\\

\textit{Weaknesses:} in ColLearning Agreement, an asymptotically optimal averaging constant remains open. It does not study the trade-off between the averaging constant that can be done and the Byzantine nodes' number that can be tolerated. Moreover, it has not applied any privacy preservation techniques.
 
%----------------------------------------------------------------------------
\subsubsection{Partially  Asynchronous training Byzantine fault tolerance approaches}
\paragraph{\textbf{Norm Filtering and Norm-cap Filtering}}

Gupta and Vaidya \cite{gupta2019byzantine_a} introduced algorithms based on deterministic gradient descent to handle the Byzantine fault tolerance problem in distributed linear regression in a multi-agent system. The authors considered a system that contained a server and many agents (workers), which might be non-faulty/faulty agents. Each agent has a fixed amount of data points and responses. The server identifies Byzantine agents using norm-based filtering techniques provided by the authors \textit{(Norm Filtering and Norm-cap Filtering)}. The proposed techniques can reinforce the distributed gradient descent algorithm deterministically when $f = n$ is less than the set threshold values. As a result, the mentioned regression problem can be solved.

The first algorithm is the \textit{Gradient Descent with Norm Filtering}: this algorithm runs in two simple steps, in which the server begins with an arbitrary estimate of the parameter and iteratively updates it.
 
\textbf{Step 1}: At the actual estimated parameter value, the server aggregates the gradients of all agent costs. It sorts them in ascending order of their $2$-norms.

\textbf{{Step 2}}: This step starts when the server uses the sum (vector) of the remaining gradients as the update direction after filtering the gradients with $f$ largest $2$-norms. Hence, the results filtering scheme is called norm filtering.

The proposed algorithms consider the partially asynchronous system, and no assumption is needed in the probability distribution of the data points. Furthermore, the computational overhead for dealing with Byzantine faulty agents is log-linear and linear in the number of agents and dimension of data points, respectively.\\

\textit{Weaknesses:} in the presented approach, only the deterministic gradient descent method was assumed.

%============================================================================
\section{Discussion and future work}
\label{DiscFutu}
%----------------------------------------------------------------------------
This paper provides the first literature review on BFT in DML. The analysis phase findings show that the proposed approaches to dealing with Byzantine faults in DML address several challenging problems; however, there are still some drawbacks. Hence, discussing the analytic findings, open issues, and research directions are critical to improving the solutions that will be addressed to BFT in DML.

\subsection{Discussion}
\label{Disc}
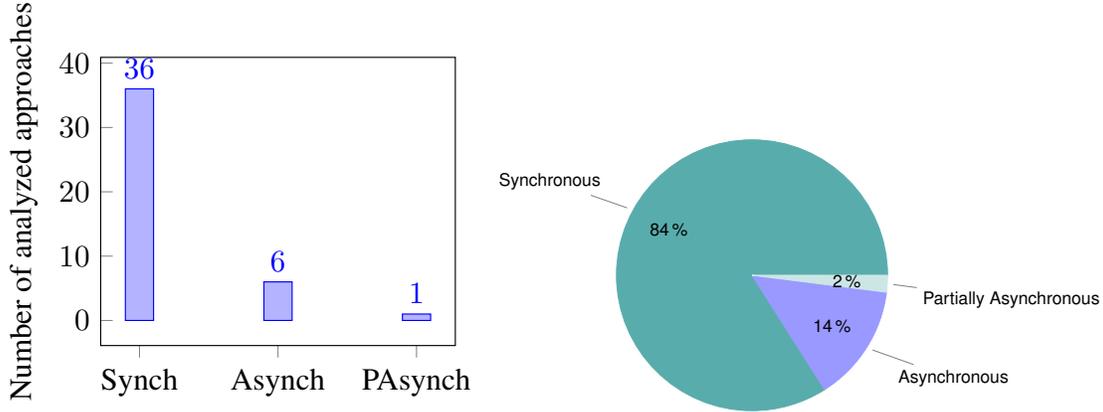
\begin{figure}[h]
  \centering
\subfloat{\resizebox{0.4\textwidth}{!}{
         \begin{tikzpicture}
\begin{axis}[
    ybar,
    enlargelimits=0.14,
    ylabel={Number of analyzed approaches},
    symbolic x coords={Synch,Asynch,PAsynch},
    xtick=data,
    nodes near coords,
    nodes near coords align={vertical},
    xtick pos=left,
    ytick pos=left
    ]
\addplot coordinates {(Synch,36) (Asynch,6) (PAsynch,1)};
\end{axis}
\end{tikzpicture}
}}
\subfloat{\resizebox{0.5\textwidth}{!}{
         \begin{tikzpicture}[nodes = {font=\sffamily}]
  \foreach \percent/\name in {
      84/Synchronous,
      14/Asynchronous,
      2/Partially Asynchronous
    } {
      \ifx\percent\empty\else               % If \percent is empty, do nothing
        \global\advance\cyclecount by 1     % Advance cyclecount
        \global\advance\ind by 1            % Advance list index
        \ifnum3<\cyclecount                 % If cyclecount is larger than list
          \global\cyclecount=0              %   reset cyclecount and
          \global\ind=0                     %   reset list index
        \fi
        \pgfmathparse{\cyclelist[\the\ind]} % Get color from cycle list
        \edef\color{\pgfmathresult}         %   and store as \color
        % Draw angle and set labels
        \draw[fill={\color},draw={\color}] (0,0) -- (\angle:\radius)
          arc (\angle:\angle+\percent*3.6:\radius) -- cycle;
        \node at (\angle+0.5*\percent*3.6:0.7*\radius) {\percent\,\%};
        \node[pin=\angle+0.5*\percent*3.6:\name]
          at (\angle+0.5*\percent*3.6:\radius) {};
        \pgfmathparse{\angle+\percent*3.6}  % Advance angle
        \xdef\angle{\pgfmathresult}         %   and store in \angle
      \fi
    };
\end{tikzpicture}   
    }}
  
  \caption{(a) Applicability count (b) Applicability percentage of Byzantine fault tolerance training processes.}
  \label{fig:AppPercBFT_TrainingProc}
\end{figure}

Analysis and evaluation are drawn in from more than 40 methods. Fig.~\ref{fig:AppPercBFT_TrainingProc} shows the applicability of the Synchronous (Synch) training process with 84\% over the Asynchronous (Asynch) training process with 14\% and 2\% of the Partially Asynchronous (PAsynch) training process \footnote{The total of the papers we analyzed is 43, divided into synchronous (36 papers), asynchronous (6 papers), and partially asynchronous (1 paper). We used the triple rule, and the result we got, we rounded it to the unit and got the ratios (84\%, 14\% and 2\%).}.  

\subsubsection{Synchronous approaches}
\label{DiscSynch}
Even though synchronous training involves waiting for slower workers, it has been used by several approaches. The latter overcame the Byzantine workers' problem under two-level settings: centralized and decentralized. 

Most centralized synchronous approaches rely largely on the assumption that honest workers lead the entire group, implying that most workers are non-Byzantine. Such approaches can therefore eliminate outliers. 

Under centralized synchronous Byzantine fault tolerance, the approaches remain relatively straightforward and can easily be implemented. Nevertheless, it is difficult to control the number of Byzantine workers. For example, when the number of Byzantine workers exceeds half, the proposed filtering schemes like Krum and Multi-Krum \cite{blanchard2017byzantine,blanchard2017machine}, DSML-Adversarial \cite{chen2017distributed}, ByzRDL \cite{yin2018byzantine}, ByzantineSGD \cite{alistarh2018byzantine},  GBT-SGD \cite{xie2018generalized} become ineffective. However, we would like to point out that the synchronous SGD method Zeno \cite{xie2018zeno} is proposed for an unlimited number of Byzantine workers.

Furthermore, analyzed filtering schemes \cite{blanchard2017byzantine, blanchard2017machine, yin2018byzantine, chen2017distributed, gupta2019byzantine_b, mhamdi2019fast} need redundant data points to ensure exact fault tolerance, which may result in increased storage and processing requirements. 

Two filtering schemes, GBT-SGD and Phocas \cite{xie2018generalized, xie2018phocas}, are the first to realize a generalized Byzantine failure with no such constraint in the synchronous SGD.

However, the filtering schemes in synchronous training still suffer from other problems, such as a single point of failure, which clearly shows in the DGDAlgorithm \cite{cao2019distributed}. 

In DGDAlgorithm, a noisy version of the genuine gradient is computed as long as the parameter server can access a small part of the dataset locally. A single point of failure can affect the DGDAlgorithm since gradients still have to be collected by a parameter server. 

Furthermore, unlike ByzRDL, the Gradient Filter CGC \cite{gupta2019byzantine_b} makes no assumptions about the probability distribution of the data points. Additionally, at the expense of fault tolerance, Gradient Filter CGC does not raise the workers' computational workload.

Most existing filtering schemes use robust aggregation rules such as geometric median and median as an alternative to simple averaging aggregation. They are used before averaging in other filtering schemes.

In the real world, the geometric median-based methods have the best accuracy performance compared to coordinate-wise median-based methods, which are usually worse in practice, and LICM-SGD \cite{yang2019byzantine} also falls into the coordinate-wise median.  

In addition, the center of attention in some works is on detecting and removing adversaries and others on detecting and rejecting outliers in gradient aggregation based on gradients passed from workers to the server, unlike SIGNSGD \cite{bernstein2018signsgd} and Stochastic-Sign SGD \cite{jin2020stochastic}  based on the sign of the gradient vectors and the stochastic sign of the gradient vectors, respectively. These methods used majority voting as a natural way to protect against less harmful defects and ensure convergence towards critical points without any guarantees on their quality. 

Moreover, these methods require the data stored in the workers are i.i.d., which is impractical in the federated learning setting. The RSA assuming aggregation model may suffer high communication costs, in contrast to Stochastic-Sign SGD, which preserves communication efficiency, can deal with heterogeneous data distribution and guarantees Byzantine resilience. Unlike RSA \cite{li2019rsa}, which does not rely on i.i.d. assumption, it uses model aggregation to find a consensus model and integrate a regularized term with the objective function. However, it needs strong convexity. 
 
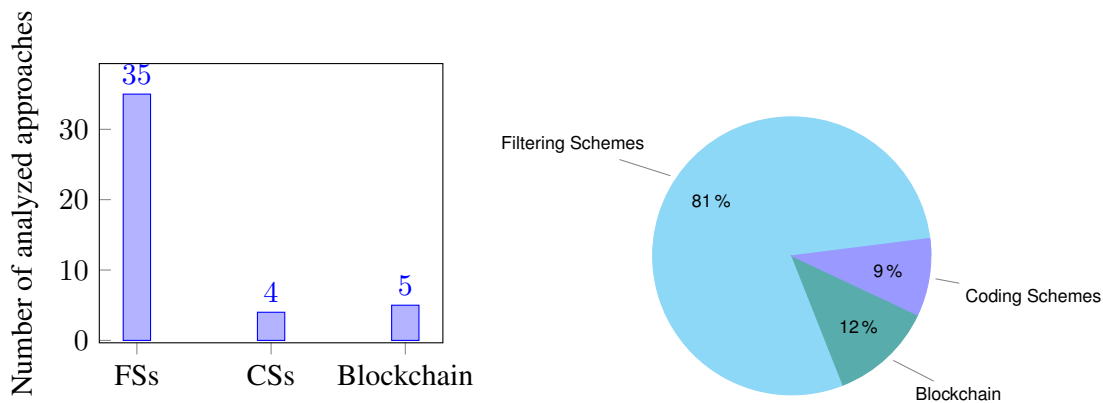
\begin{figure}[h]
  \centering
  \subfloat{\resizebox{0.4\textwidth}{!}{
         \begin{tikzpicture}
\begin{axis}[
    ybar,
    enlargelimits=0.14,
    ylabel={Number of analyzed approaches},
    symbolic x coords={FSs,CSs,Blockchain},
    xtick=data,
    nodes near coords,
    nodes near coords align={vertical},
     xtick pos=left,
    ytick pos=left
    ]
\addplot coordinates {(FSs,35) (CSs,4) (Blockchain,5)};
\end{axis}
\end{tikzpicture}
}}
\subfloat{\resizebox{0.5\textwidth}{!}{
         \begin{tikzpicture}[nodes = {font=\sffamily}]
  \foreach \percent/\name in {
      81/Filtering Schemes,
      12/Blockchain,
      9/Coding Schemes
    } {
      \ifx\percent\empty\else               % If \percent is empty, do nothing
        \global\advance\cyclecount by 1     % Advance cyclecount
        \global\advance\ind by 1            % Advance list index
        \ifnum3<\cyclecount                 % If cyclecount is larger than list
          \global\cyclecount=0              %   reset cyclecount and
          \global\ind=0                     %   reset list index
        \fi
        \pgfmathparse{\cyclelist[\the\ind]} % Get color from cycle list
        \edef\color{\pgfmathresult}         %   and store as \color
        % Draw angle and set labels
        \draw[fill={\color},draw={\color}] (0,0) -- (\angle:\radius)
          arc (\angle:\angle+\percent*3.6:\radius) -- cycle;
        \node at (\angle+0.5*\percent*3.6:0.7*\radius) {\percent\,\%};
        \node[pin=\angle+0.5*\percent*3.6:\name]
          at (\angle+0.5*\percent*3.6:\radius) {};
        \pgfmathparse{\angle+\percent*3.6}  % Advance angle
        \xdef\angle{\pgfmathresult}         %   and store in \angle
      \fi
    };
\end{tikzpicture}
}}
  \caption{(a) Applicability count (b) Applicability percentage of Byzantine fault tolerance schemes.}
  \label{fig:AppPercBFT_schemes}
\end{figure}

The applicability of the Filtering Schemes (FSs) represents the most existence-based methods by 81\%, contrary to Coding Schemes (CSs), which are represented by 9\%, as shown in Fig.~\ref{fig:AppPercBFT_schemes}.

DRACO \cite{chen2018draco} is a coding scheme that guarantees correct gradients recovery by increasing the workers' computational workload. Unlike Bulyan \cite{el2018hidden}, which requires heavy computational overhead on the parameter server side with $4f + 3$ workers, DRACO requires only $2f +1$ workers, making it strongly Byzantine-resilient. However, the computation time of DRACO may be many times due to the encoding and decoding time, accordingly larger than the computation time of ordinary SGD, and the proposed method in AGGREGATHOR \cite{damaskinos2019aggregathor} with the redundant calculations can avoid this overhead. In addition, the storage redundancy factor required in DRACO compared to BRDO-based data encoding \cite{data2020data} increases linearly with the number of corrupted worker nodes, unlike BRDO-based data encoding, which is constant, which results in low compute at the worker node level. The authors in DETOX \cite{rajput2019detox} improve resiliency and overhead guarantees by combining a deterministic coding scheme with a robust aggregation rule; also, they show that computational efficiency will be enhanced in DRACO at the expense of exact fault tolerance when using the gradient filters.

The above coding schemes based methods (DRACO, BRDO-based data encoding, DETOX) are so close to each other that they use a similar deterministic coding scheme. However, BRDO-based Data encoding has a $2n/(n-2f)$ storage redundancy factor, in contrast to DRACO, which has a $2f +1$ storage redundancy factor, and with the corrupt workers' number increases linearly. As far as we know, RRR-BFT \cite{gupta2019randomized} is the first proposed method that tolerates Byzantine workers in the context of parallel learning by using the idea of reactive redundancy. In addition, RRR-BFT compares the computational efficiency of its deterministic coding schemes to DRACO and enhances it using the randomization technique. Furthermore, the server in the synchronous setting can improve the computing efficiency by combining the gradient-filter method with the randomized or deterministic coding scheme in RRR-BFT or DETOX, respectively. However, according to the authors, when the server employed the gradient filter, it could not identify all of the existing Byzantine workers. The most current methods in coding schemes are based on deterministic coding schemes such as DRACO, DETOX, and BRDO data encoding. At the same time, RRT-BFT used two types of coding schemes (deterministic and randomized).

The most proposed centralized synchronous approaches are based on the SGD first-order optimization method. However,  other publications, such as Distributed Momentum \cite{el2020distributed}, DRSL-Byzantine Mirror Descent \cite{ding2019distributed}, and SVRG-AdversarialML \cite{bulusu2019convex}, explore Momentum, Mirror Descent, and SVRG, respectively.  Momentum reduces gradient variation at the server and reinforces Byzantine resilient aggregation rules. Mirror Descent is used to secure training data against faulty information sharing. In SVRG-AdversarialML, the authors benefit from the advantages of SVRG over SGD to reduce variance and propose the first method-based SVRG to fight against Byzantine adversaries in the distributed setting.

The main drawback of the discussed synchronous centralized methods is that they keep the training convergence and correctness at the expense of useful information. Also, these methods require assumptions that are not easy to apply in the federated learning paradigm, such as the need to relocate data points in DRACO. Additionally, the i.i.d. assumption is not the case in federated learning, where the computing units are heterogeneous, and there are difficulties in generalizing the existing algorithms in the non-i.i.d. setting. Furthermore, the existence of heterogeneous workers will certainly be due to the slowness of convergence. Finally, synchronizing with federated learning and edge computing offline workers cannot be done most of the time.

On the other hand, decentralized synchronous methods are recently represented by multi-server parameter method as (GuanYu \cite{el2019sgd}, DSML-Adversarial and Securing-DML \cite{su2018securing}), without any server as (ByRDiE \cite{yang2019byrdie}, BRIDGE \cite{yang2019bridge} and MOZI \cite{guo2020towards}) and with Blockchain technology as (DeepChain \cite{weng2019deepchain}, LearningChain \cite{chen2018machine}, TCLearn \cite{lugan2019secure}, SDPP based-Blockchain System \cite{zhao2019mobile} and BlockDeepNet \cite{rathore2019blockdeepnet}).

To simplify the explanation for the proposed DSML-Adversarial, the authors assume in their article that there is only one parameter server, which is why we refer to the presented DSML-Adversarial with centralized synchronous methods discussed above. In addition, they show the applicability of their proposed method for multi-server parameters through their algorithm description and detailed analysis. GuanYu and Securing-DML assume the same decentralized setting as DSML-Adversarial with multi-server parameters. In addition, the parameter server architecture used in the centralized setting can be vulnerable to a single failure point, as in DGDAlgorithm. So, eliminating the parameter server is necessary to avoid this issue and get a fully decentralized setting. 

In ByRDiE and BRIDGE, the authors assume a fully decentralized setting. ByRDiE applies the trimmed-median rule to the coordinate descent optimization algorithm, and BRIDGE applies it to the SGD algorithm. Nevertheless, BRIDGE has the best communication cost compared to ByRDiE. However, both remain vulnerable to some Byzantine attacks, in contrast to Mozi, which deals with simple and sophisticated attacks with low computation overhead. Likewise, we survey several decentralized methods (DeepChain, LearningChain, TCLearn, SDPP-based-Blockchain System, BlockDeepNet) based on Blockchain technology. These methods mitigate the challenges in previous works, such as poisoning attacks, single point of failure, data privacy, and existing heterogeneous workers as federated learning. Most Blockchain-based methods achieve collaborative training and improve security, privacy, and auditability.

Through Fig.~\ref{fig:AppPercBFT_TrainingProc}, we notice that 12\% of methods used Blockchain technology, which shows that researchers are slightly using them compared to filtering schemes.

In the synchronous settings, the authors can assume that the network is perfect and delivers messages within a specific time, which remains relatively simple and easy to be implemented, so we find that most of the proposed approaches considered the synchronous setting by 84\% compared to 14\% of the asynchronous approaches. Asynchronous settings are more practical and realistic since no network can be assumed to be perfect and require more complex implementation.

\subsubsection{Asynchronous approaches}
\label{DiscASynch}
Asynchronous Byzantine learning is more general than synchronous Byzantine learning, which motivates authors to change the research direction to the asynchronous setting. In addition, it is necessary for workers in the synchronous setting with better computing capacity to wait for the struggling workers, leading to wasted computing resources. As a result, asynchronous Byzantine learning is more convenient than synchronous Byzantine learning, which is another motivation for the authors to propose solutions in asynchronous settings to solve Byzantine learning problems.

Kardam \cite{damaskinos2018asynchronous} is the first proposed method that tolerates Byzantine learning in the asynchronous distributed SGD algorithm. Kardam filters out the outliers using a different filtering scheme method than those used in the synchronous approaches. Its filtering scheme is based on the Lipschitzness filter and frequency filter. However, Kardam cannot resist malicious attacks due to the weak threat model assumption. 

The second proposed method to deal with Byzantine attackers in the asynchronous distributed SGD algorithm is proposed in DBT-SGD in the Era of Big Data \cite{jin2019distributed}. Nonetheless, the authors proposed two different algorithms in a decentralized setting. They avoid the problem of a single point of failure caused by the parameter server and repose to the collaborative training among the set of workers. To the best of our knowledge, DBT-SGD in the Era of Big Data is the first paper proposed to resolve the Byzantine problems in the asynchronous decentralized setting. However, it can also be implemented with centralized settings, which makes it vulnerable to a single point of failure. 

Another work based on the asynchronous setting is Zeno++ \cite{xie2020zeno++}. The author in Zeno++ proposes a strong threat model compared to the Kardam one; it also proves many advantages, unlike Kardam, which needs a majority of honest workers and workers' bounded staleness. However, Zeno++ stores training data on the server for scoring will increase the risk of confidentiality leakage. 

BASGD \cite{yang2020basgd} uses two essential components (buffer and aggregation function), proving its ability to resist error and malicious attack compared to Kardam. Unlike Zeno++, it avoids the problem of storing data on the server and preserves privacy in distributed learning.   

On the other hand, To the best of our knowledge, ByzSGD \cite{el2020genuinely} and  ColLearning Agreement \cite{ el2020collaborative} are the only proposed approaches in the decentralized asynchronous settings with DBT-SGD in the Era of Big Data. ByzSGD considers the decentralized parameter server settings and increases the efforts made by the centralized synchronous algorithms, as a result tolerating Byzantine servers and Byzantine workers. Nevertheless, ColLearning Agreement considers the fully decentralized settings as ByRDiE, BRIDGE, and MOZI. However, these methods suppose that the data is i.i.d. and consider only convex optimization. In genuine Byzantine environments, the correct models can be influenced by Byzantine workers to distance them from each other. In comparison, MOZI assumes that the models do not drift from each other on good workers, which is impractical. Moreover, despite the two ByzSGD hypotheses (configuration of decentralized parameter servers and genuinely Byzantine adversaries) with the assumption that the workers are drawn from the homogeneous, i.i.d., this solution is not collaborative like ColLearning Agreement.

\subsubsection{Partially asynchronous approaches}
\label{DiscPAsynch}
To the best of our knowledge, the partially asynchronous training Byzantine fault tolerance setting is used by only one work \cite{gupta2019byzantine_a}. It takes advantage of this setting regarding system robustness to bounded delay and achieves a log-linear and linear computational overhead.

\begin{figure}[h]
  \centering
  \subfloat{\resizebox{0.4\textwidth}{!}{
         \begin{tikzpicture} 
\begin{axis}[
    ybar,
    enlargelimits=0.14,
    legend style={at={(0.5,-0.18)},
      anchor=north,legend columns=-1},
    ylabel={Number of analyzed approaches},
    symbolic x coords={Synch,Asynch,PAsynch},
    xtick=data,
    nodes near coords,
    nodes near coords align={vertical},
     xtick pos=left,
    ytick pos=left
    ]
\addplot coordinates {(Synch,25) (Asynch,3) (PAsynch,1)};
\addplot coordinates {(Synch,11) (Asynch,3) (PAsynch,0)};
\legend{Centralized,Decentralized}
\end{axis}
\end{tikzpicture}
}}
\subfloat{\resizebox{0.5\textwidth}{!}{
         \begin{tikzpicture}[nodes = {font=\sffamily}]
  \foreach \percent/\name in {
      58/Synch Centralized,
      26/Synch Decentralized,
      7/Asynch Centralized,
      7/Asynch Decentralized,
      2/PAsynch Centralized,
    } {
      \ifx\percent\empty\else               % If \percent is empty, do nothing
        \global\advance\cyclecount by 1     % Advance cyclecount
        \global\advance\ind by 1            % Advance list index
        \ifnum3<\cyclecount                 % If cyclecount is larger than list
          \global\cyclecount=0              %   reset cyclecount and
          \global\ind=0                     %   reset list index
        \fi
        \pgfmathparse{\cyclelist[\the\ind]} % Get color from cycle list
        \edef\color{\pgfmathresult}         %   and store as \color
        % Draw angle and set labels
        \draw[fill={\color},draw={\color}] (0,0) -- (\angle:\radius)
          arc (\angle:\angle+\percent*3.6:\radius) -- cycle;
        \node at (\angle+0.5*\percent*3.6:0.7*\radius) {\percent\,\%};
        \node[pin=\angle+0.5*\percent*3.6:\name]
          at (\angle+0.5*\percent*3.6:\radius) {};
        \pgfmathparse{\angle+\percent*3.6}  % Advance angle
        \xdef\angle{\pgfmathresult}         %   and store in \angle
      \fi
    };
\end{tikzpicture}
}}
  \caption{(a) Applicability count (b) Applicability percentage of Byzantine fault tolerance training processes and settings.}
  \label{fig:AppPercBFT_TrainingProcandSett}
\end{figure}
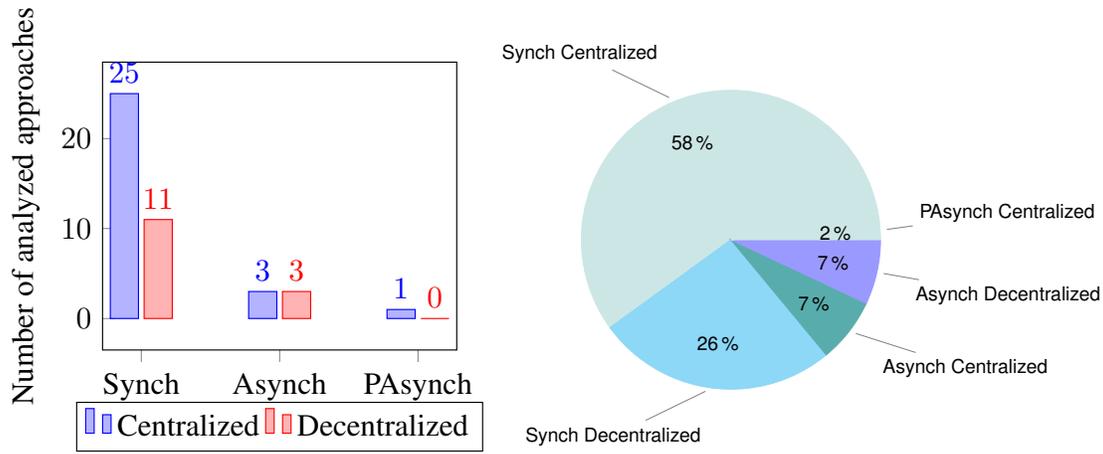

The distribution of the applicability of the Byzantine fault tolerance training processes and settings by discussed methods is provided in Fig.~\ref{fig:AppPercBFT_TrainingProcandSett}.

Despite the fact that Byzantine fault tolerance approaches are extensively used, research continues. The ongoing advancements in distributed machine learning, particularly federated learning, and the integration with other techniques, such as IoT or fog computing, increase systems' complexity and the probability of failure. As a result, the area of study on Byzantine fault tolerance strategies in distributed machine learning will be expanded.

The BFT field aims to preserve the system's continuity and dependability during arbitrary behavior. BFT is extremely important as it keeps the system operating even if certain components fail. Many machine learning algorithms have an objective function that expresses their goal. Commonly, the learning algorithm minimizes the objective function to optimize the model. For example, a low error in the case of classifying customer satisfaction score in e-commerce into positive or negative, or incoming messages in the mail into legitimate or spam. The training process is terminated when a near-optimal/optimal solution is identified, or the model is converged. Byzantine refers to the malicious nodes (workers/servers) and their (sent/update) gradients. For example, the Byzantine workers lead the objective function to converge to ineffective models and preclude the distributed algorithm process from converging to a satisfactory state. The latter is the state that accomplishes an accuracy with Byzantine workers equivalent to the one reached without any Byzantine workers. Several BFT solutions in DML have achieved good results in terms of convergence, time complexity, and accuracy.

A summary of analysis methods under several metrics (type of scheme, type of optimization method, centralized and decentralized settings) are given in Table 4, Table 5, Table 6, and Table 7. We point out that ( $ \bullet $ ) means applicable, (N.A) means not applicable, and ($-$) indicates that, to the best of our knowledge, information is not supplied in the original resource or other studies.

\begin{table} [H]
  \caption{Summarizing different Synchronous Byzantine fault tolerance approaches.}
  \label{tab:Syncomp}
  \begin{minipage}{\columnwidth}
  \begin{center}
  \begin{tabular}{p{3.9 cm} c c c c p{2 cm}}

\toprule
    Approach&Filtering scheme&Coding scheme &Blockchain&Optimization method\\
    \midrule
    DeepChain& N.A& N.A&$ \bullet $&SGD\\
    Krum, Multi-Krum&$ \bullet $&N.A &N.A &SGD\\
    DSML-Adversarial&$ \bullet $& N.A& N.A&GD\\
    ByzRDL&$ \bullet $&N.A &N.A &GD\\    
    ByzantineSGD&$ \bullet $&N.A&N.A&SGD\\
    GBT-SGD&$ \bullet $&N.A&N.A&SGD\\
    DGDAlgorithm&$ \bullet $&N.A&N.A&GD\\
    Securing-DML&$ \bullet $&N.A&N.A&GD\\
    Tremean and Phocas&$ \bullet $&N.A&N.A&SGD\\
    DRACO\footnote{applicable also to any first order method.}&N.A &$ \bullet $&N.A &Mini-batch SGD\\
    Bulyan&$ \bullet $&N.A&N.A&SGD\\
    ByzantinePGD\footnote{PGD:Perturbed Gradient Descent}&$ \bullet $&N.A&N.A&GD, PGD\\
    Zeno&$ \bullet $&N.A&N.A&SGD\\
    LearningChain&N.A&N.A&$ \bullet $&SGD\footnote{LearningChainEx applied to any gradient descent based machine learning systems.}\\
    RSA & $ \bullet $ &N.A&N.A& SGD\\
    signSGD & $ \bullet $ &N.A&N.A& SGD\\
    BRDO based Data Encoding\footnote{PGD: Proximal Gradient Descent, CD: Coordinate Descent}&N.A&$ \bullet $&N.A&PGD and CD\\
    AGGREGATHOR&$ \bullet $&N.A&N.A&SGD\\
    GuanYu&$ \bullet $&N.A&N.A&SGD\\
    Trimmed mean&$ \bullet $&N.A&N.A&SGD\\
    TCLearn&N.A&N.A&$ \bullet $&GD\\
    SDPP based-Blockchain System&N.A&N.A&$ \bullet $&$-$\\
    ByRDiE&$ \bullet $&N.A&N.A&Coordinate Descent\\
    BlockDeepNet&N.A&N.A&$ \bullet $&GD\\
    FABA&$ \bullet $&N.A&N.A&SGD\\
    BRIDGE&$ \bullet $&N.A&N.A&SGD\\
    LICM-SGD&$ \bullet $&N.A&N.A&SGD\\
    Gradient-Filter CGC & $ \bullet $&N.A &N.A & SGD\\
    SVRG-AdversarialML&$ \bullet $&N.A&N.A&SVRG\\
    LIUBEI&$ \bullet $&N.A&N.A&SGD\\
    DRSL-Byzantine Mirror Descent&$ \bullet $&N.A&N.A&Mirror Descent\\
    DETOX&$ \bullet $&$ \bullet $&N.A&Mini-batch SGD\\
    RRR-BFT &N.A&$ \bullet $&N.A & SGD\\
    MOZI&$ \bullet $&N.A&N.A&$-$\\ 
    Stochastic-Sign SGD &$ \bullet $& N.A& N.A& SGD\\ 
    Distributed Momentum&$ \bullet $&N.A &N.A &Momentum\\  
  \bottomrule
\end{tabular}
\end{center}
\bigskip
\end{minipage}
\end{table}

\begin{table}[H]
  \caption{Summarizing different Asynchronous Byzantine fault tolerance approaches}
  \label{tab:ASyncomp}
  \begin{minipage}{\columnwidth}
  \begin{center}
\begin{tabular}{p{4 cm} c p{3 cm} c p{3 cm} c p{2 cm}p{2 cm}}

\toprule
    Approach&Filtering scheme&Coding scheme &Blockchain&Optimization method\\
    \midrule
    Kardam&$ \bullet $&N.A&N.A&SGD\\
    DBT-SGD in the Era of Big Data&$ \bullet $&N.A&N.A&SGD\\
    Zeno++&$ \bullet $&N.A&N.A&SGD\\
    BASGD&$ \bullet $&N.A&N.A&SGD\\
    ByzSGD&$ \bullet $&N.A&N.A&SGD\\
    ColLearning Agreement &$ \bullet $&N.A&N.A&SGD\\
   
  \bottomrule
\end{tabular}
\end{center}
\bigskip
\end{minipage}
\end{table}
\begin{table}[H]
  \caption{Summarizing different Partially Asynchronous Byzantine fault tolerance approaches}
  \label{tab:PASynch}
  \begin{minipage}{\columnwidth}
  \begin{center}
 \begin{tabular}{p{3 cm} c p{3 cm} c p{3 cm} c p{2 cm}p{2 cm}}

\toprule
    Approach&Filtering scheme&Coding scheme &Blockchain&Optimization method\\
    \midrule
    
    Norm Filtering and Norm-cap Filtering&$ \bullet $&N.A&N.A&GD\\
\bottomrule
\end{tabular}
\end{center}
\bigskip
\end{minipage}
\end{table}

\begin{table}
  \caption{Summarizing different Centralized/Decentralized Byzantine fault tolerance approaches}
  \label{tab:CenDec}
  \begin{minipage}{\columnwidth}
  \begin{center}
  \begin{tabular}{p{4 cm} p{5 cm} p{5 cm}}

\toprule
Training process&Centralized&Decentralized\\
    \midrule
Synchronous&Krum and Multi-Krum \cite{blanchard2017byzantine,blanchard2017machine}, 
ByzRDL\cite{yin2018byzantine},
ByzantineSGD \cite{alistarh2018byzantine},
GBT-SGD \cite{xie2018generalized},
DGDAlgorithm \cite{cao2019distributed},
Phocas \cite{xie2018phocas},
DRACO \cite{chen2018draco},
Bulyan \cite{el2018hidden},
ByzantinePGD \cite{yin2019defending},
Zeno \cite{xie2018zeno},
RSA \cite{li2019rsa},
SIGNSGD \cite{bernstein2018signsgd},
BRDO based Data Encoding \cite{data2020data},
AGGREGATHOR \cite{damaskinos2019aggregathor},
Trimmed mean \cite{ tianxiang2019aggregation},
FABA \cite{xia2019faba},
LICM-SGD \cite{yang2019byzantine},
Gradient Filter CGC \cite{gupta2019byzantine_b},
SVRG-AdversarialML \cite{bulusu2019convex},
LIUBEI \cite{mhamdi2019fast},
DRSL-Byzantine Mirror Descent \cite{ding2019distributed},
DETOX \cite{rajput2019detox},
RRR-BFT \cite{gupta2019randomized},
Stochastic-Sign SGD \cite{jin2020stochastic}, 
Distributed Momentum \cite{el2020distributed}& DeepChain \cite{weng2019deepchain},
DSML-Adversarial\cite{chen2017distributed},
Securing-DML \cite{su2018securing},
GuanYu \cite{el2019sgd},
ByRDiE \cite{yang2019byrdie},
BRIDGE \cite{yang2019bridge},
MOZI \cite{guo2020towards},
LearningChain \cite{chen2018machine},
TCLearn \cite{lugan2019secure},
SDPP based-Blockchain System \cite{zhao2019mobile},
BlockDeepNet \cite{rathore2019blockdeepnet}\\

\\Asynchronous&Kardam \cite{damaskinos2018asynchronous},
Zeno++ \cite{xie2020zeno++},
BASGD \cite{yang2020basgd}&DBT-SGD in the Era of Big Data \cite{jin2019distributed},
ByzSGD \cite{el2020genuinely},
ColLearning Agreement \cite{ el2020collaborative} \\
\\Partially Asynchronous&Norm Filtering and Norm-cap Filtering\cite{gupta2019byzantine_a}&N.A\\
	
\bottomrule
\end{tabular}
\end{center}
\bigskip
\end{minipage}
\end{table}
\subsection{Potential future direction}
\label{Futu}
Based on the discussion above and for future research, the following directions areas are recommended:
\begin{enumerate}
\item The most used optimization methods for the BFT problem in DML belong to the first-order optimization methods. These methods achieved important results, and some challenges remain unresolved, such as high computational complexity and data privacy. Second-order methods are an interesting family optimization method that may produce more effective outcomes than first-order methods.
\item SGD is the popular method used in the BFT problem in DML, while it presents challenges that have led to its improvement through proposed variants like (Adagrad, Adadelta, RMSprop ... etc.). That implies that the existing BFT approaches in DML with SGD may be improved using these variants instead of the SGD method. As a result, it is interesting coming up with new solutions with these variants to create a more robust, reliable, and secure DML.
\item Most of the analyzed approaches are based on synchronous/asynchronous training processes in centralized/decentralized settings (fewer approaches in the asynchronous). Overdone asynchronism can be hurtful to the convergence of some algorithms \cite{bertsekas1989parallel}, and synchronization cannot be done most of the time (case of federated learning). It would be exciting to enrich the results obtained by the analyzed approaches using other training processes, such as (semi-synchronous or partially asynchronous). In other words, exploring existing training processes can be used to avoid some problems in synchronous/asynchronous training processes.
\item The widely used IoT means more devices, more data, and more requirements for storing, transmitting, and processing that data, as well as keeping it secure. The emergence of new paradigms (IoT, big data, edge computing, etc.) implicates several challenges, including ensuring the same or better results regardless of scale participant number and ideally decreasing cost communication. Blockchain technology showed promising results in the BFT problem in DML concerning data confidentiality and ensuring trusted collaborative model training(considering that Blockchain was used sparingly). Thus, Blockchain is an interesting technology for exploring new solutions to open problems in current paradigms.
\end{enumerate}

\begin{figure*}
    \centering
    \includegraphics[width=\textwidth,keepaspectratio]{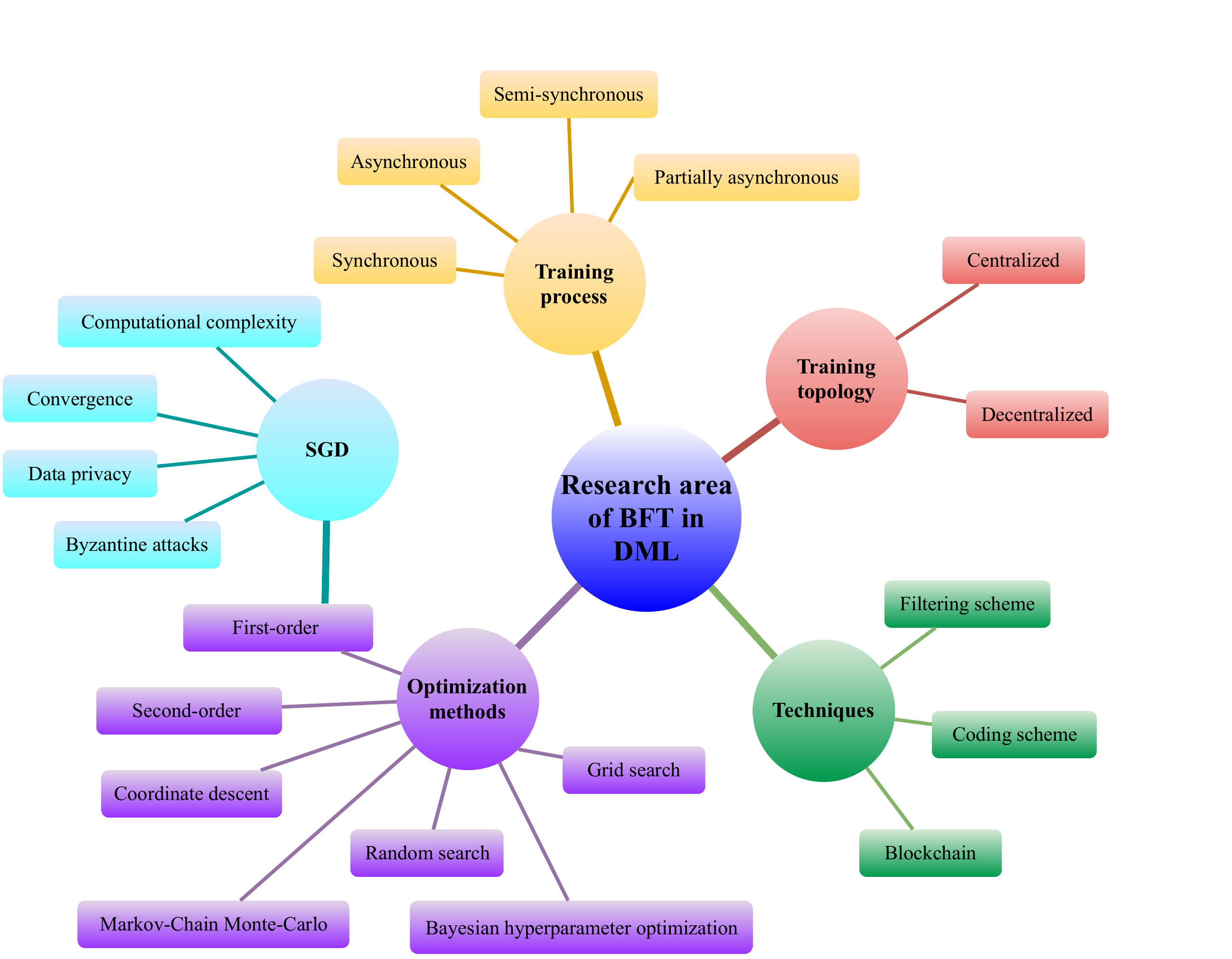}
    \caption{Mind map of the BFT research area in DML. The mind map depicts the different optimization method families that can be used: existing techniques that may be developing (such as filtering scheme, coding scheme and Blockchain), training topologies (as centralized and decentralized), training processes that can be combined in various ways, and existing problems in SGD, the most commonly used optimization method in DML, to propose more effective solutions.
}
    \label{fig:BFTinDML}
\end{figure*}

The possibilities of the research area are summarized in Fig.~\ref{fig:BFTinDML}.

%============================================================================
\section{Conclusion}
\label{Conclusion}
%----------------------------------------------------------------------------

The Byzantine fault tolerance of a distributed machine learning system is a characteristic that makes the system more reliable. Preventing the model training convergence by Byzantine workers in distributed machine learning systems is a significant issue. The first-order optimization methods are improved to resolve this type of problem. In this paper, we provided a general survey of recent Byzantine fault tolerance approaches, which have been discussed, and the following conclusions have been made:
\begin{itemize}
\item In most analyzed works, synchronous settings are favored over asynchronous ones and centralized settings over decentralized ones.
\item The filtering scheme is used more frequently than the coding scheme and Blockchain technology in the studied works.
\item In spite of the limited number of Blockchain-based approaches proposed these approaches have achieved several important outcomes (avoiding a single point of failure, achieving collaborative learning and data privacy, and preventing poisoning attacks) compared to other schemes. Blockchain can enhance previous findings and broaden its applicability in many settings by combining multiple schemes.
\item Developing a resilient distributed algorithm in DML can be difficult when nodes fail, or specific nodes behave adversarially; hence, the issue of BFT in DML has grown in importance. Mainly, we discussed that handling this challenging problem requires dealing with node failures, heterogeneous nodes, malicious attackers, topology settings, and communication issues. In addition, it is necessary to consider DML progress, such as federated learning to deal with the above difficulties. The increased data collection and technological advancements have made federated learning an increasingly popular concept in recent years. However, a substantial number of participating nodes with multiple data sources makes federated learning vulnerable to adversaries \cite{liu2021redundancy}. Moreover, nodes participating in federated learning may be unstable because they typically rely on weaker communication mediums and battery-powered computers, which tend to fail more often and drop out\cite{kairouz2021advances}. As a result, a Byzantine fault tolerance in federated learning complements this survey and represents our future work.
\end{itemize}

This paper includes a comparative study of the discussed approaches and shows possible future directions for research in the context of Byzantine fault tolerance in distributed machine learning systems.
%============================================================================
\section*{Acknowledgments}
The authors would like to thank Yuan Yin from Sorbonne University, CNRS, ISIR, F-75005 Paris, France, for insightful remarks and helpful input. The authors would also like to thank Professor Kamal Eddine Melkemi from the Department of Computer Science at Batna 2 University for useful discussion on machine learning and deep learning.

%============================================================================
%Where the bibliography will be printed
\printbibliography
%%%%
%\bibliographystyle{IEEEtran}
%\bibliography{IEEEabrv,sample-base}
\end{document}